\documentclass[final,1p,times,twocolumn]{elsarticle}
\usepackage{amsmath}
\usepackage[T1]{fontenc}
\usepackage[dvipdfm,colorlinks=true,citecolor=blue,linkcolor=blue,urlcolor=blue]{hyperref}
\everymath{\displaystyle}

\usepackage{lineno}

\journal{Annals of Nuclear Energy}

\begin{document}

\begin{frontmatter}

\title{Evaluation of uranium-233 neutron capture cross section in keV region}

\author[IAEA]{Naohiko Otuka}
\ead{n.otsuka@iaea.org}
\affiliation[IAEA]{organization={Nuclear Data Section, Division of Physical and Chemical Sciences, Department of Nuclear Sciences and Applications, International Atomic Energy Agency},
            addressline={Wagramerstra\ss e 5}, 
            city={A-1400 Wien},
            country={Austria}}

\author[JAEA1]{Kenichi Tada}
\ead{tada.kenichi@jaea.go.jp}
\affiliation[JAEA1]{organization={Research Group for Reactor Physics and Thermal Hydraulics Technology, Japan Atomic Energy Agency},
            addressline={Tokai-mura, Naka-gun}, 
            city={Ibaraki},
            postcode={319-1195},
            country={Japan}}

\author[UPM1,UPM2]{Oscar Cabellos}
\ead{oscar.cabellos@upm.es}
\affiliation[UPM1]{organization={Department of Energy Engineering (Division of Nuclear Engineering), Universidad Polit\'{e}cnica de Madrid},
             city={28006 Madrid},
             country={Spain}}
\affiliation[UPM2]{organization={Instituto de Fusi\'{o}n Nuclear - ``Guillermo Velarde", Universidad Polit\'{e}cnica de Madrid},
             city={28006 Madrid},
             country={Spain}}

\author[JAEA2]{Osamu Iwamoto}
\ead{iwamoto.osamu@jaea.go.jp}
\affiliation[JAEA2]{organization={Nuclear Data Center, Japan Atomic Energy Agency},
            addressline={Tokai-mura, Naka-gun}, 
            city={Ibaraki},
            postcode={319-1195},
            country={Japan}}

\begin{abstract}
The uranium-233 neutron capture cross section between 3~keV and 1~MeV was evaluated with the new alpha value recently measured at the Los Alamos National Laboratory LANCE facility and compiled in the EXFOR library.
The obtained capture cross section is systematically lower than those in the latest versions of the major general purpose nuclear data libraries, and the reduction from the JENDL-5 library is close to 50\% around 20~keV.
The newly evaluated cross section was benchmarked against 166 criticality experiments chosen from the ICSBEP handbook by performing Monte Carlo neutron transport calculation with the JENDL-5 library,
and slight reduction of the cumulative chi-square value was achieved by adoption of the newly evaluated capture cross section.
\end{abstract}


\end{frontmatter}

\section{Introduction}
The Th-U fuel cycle has been an attractive alternative to the U-Pu fuel cycle because of the higher thermal neutron capture cross section of $^{232}$Th and the higher neutron reproduction factor (eta value) of $^{233}$U~\cite{IAEA2005Thorium}.
As the accuracy of the nuclear data required for designs of the Th-U cycle related reactor systems was not satisfactory,
IAEA Consultants' Meeting on assessment of nuclear data needs for thorium and other advanced cycles~\cite{Pronyaev1999Summary} determined the target accuracies of the nuclear reaction data for the Th-U systems with thermal reactors, fast reactors and accelerator-driven subcritical systems.
The meeting was followed by an IAEA Coordination Research Project on ``Evaluated nuclear data for nuclides within the thorium-uranium fuel cycle",
which released a nuclear reaction data library of $^{232}$Th, $^{231,233}$Pa and $^{232,233,234,236}$U~\cite{IAEA2010Evaluated}.
Among the cross sections of these nuclides,
$^{233}$U(n,$\gamma$)$^{234}$U cross section has been identified as one still requiring improvement and listed in the General Request (GR) category of the NEA High Priority Request list (HPRL)~\cite{Dupont2020HPRL}.
For this reaction,
the IAEA Consultants' Meeting set the target accuracy to 3\% for the fast reactor and accelerator-driven neutron fields, while the NEA HPRL sets it to 9\% between 10~keV and 1~MeV.
However, the difference of the $^{233}$U(n,$\gamma$)$^{234}$U cross sections between the latest versions of the major general purpose libraries (CENDL-3.2~\cite{Ge2020CENDL}, ENDF/B-VIII.0~\cite{Brown2018ENDF}, JEFF-3.3~\cite{Plompen2020Joint} and JENDL-5~\cite{Iwamoto2023Japanese}) exceeds the target accuracies.

Experimental determination of the capture cross section is challenging.
Identification of capture events by prompt gamma detection is not straightforward since more prompt gammas are from fission events.
The capture product $^{234}$U has extremely long half-life (245.5~ky~\cite{Kondev2021NUBASE2020}) and the conventional activation method  is not applicable to determination of the capture cross section.
Above 1~keV, the measurement of the capture-to-fission cross section ratio (alpha value) performed by Hopkins et al. in 1960s~\cite{Hopkins1962Neutron} between 30~keV and 1~MeV was the only experimental dataset providing the energy dependent capture cross sections in the EXFOR library~\cite{Otuka2014Towards},
and more experimental constraint has been desired.

Recently, a new measurement of the alpha value was performed by Leal-Cidoncha et al. at the Los Alamos Neutron Science Center (LANSCE)~\cite{Leal-Cidoncha2023Measurement}.
Among the recent neutron capture measurements of $^{233,235}$U and $^{239}$Pu with DANCE (Device for Advanced Neutron Capture Experiments),
the $^{233}$U measurement result is available in the EXFOR library with point-wise partial uncertainties.
Furthermore,
it is compiled in the EXFOR library in the form of the alpha value, which is the quantity what the experiment directly measured.
Use of an experimental alpha value converted to the corresponding absolute capture cross section may introduce bias in the evaluation,
and adoption of the alpha value itself would make evaluations more adequate than adoption of the alpha value converted to the capture cross section as long as enough information on the fission cross section is provided to the evaluation framework.

We developed an experimental database of the $^{233,235,238}$U and $^{239,240,241}$Pu fast neutron fission cross sections and their ratios for a simultaneous evaluation~\cite{Otuka2022aEXFOR}.
The $^{233}$U(n,f) cross section above 10~keV from this simultaneous evaluation is adopted in the JENDL-5 library without any adjustment,
and the CERN n\_TOF collaboration confirmed that their new $^{233}$U(n,f)/$^{235}$U(n,f) ratio between 10~keV and 1~MeV agrees with the ratio from our simultaneous evaluation within 0.5\%~\cite{Tarrio2023Neutron}.
This experimental database was also successfully applied to evaluation of the $^{232}$Th(n,f) cross section between 500~keV and 200~MeV~\cite{Devi2024EXFOR} and the $^{242}$Pu(n,f) cross section between 100~keV and 200~MeV~\cite{Okuyama2024EXFOR}.
Therefore, the experimental database originally developed for evaluation of the $^{233,235,238}$U and $^{239,240,241}$Pu(n,f) cross sections can be also a solid base for evaluation of the $^{233}$U(n,$\gamma$)$^{234}$U cross section.

The purpose of the present work is to evaluate the $^{233}$U(n,$\gamma$)$^{234}$U cross section between 3~keV and 1~MeV in the simultaneous evaluation framework by considering the new experimental $^{233}$U alpha value,
and to see impacts of the new evaluation on integral benchmarks.

\section{Evaluations in major libraries}
\begin{figure}[hbtp]
\includegraphics[angle=-90,width=\linewidth]{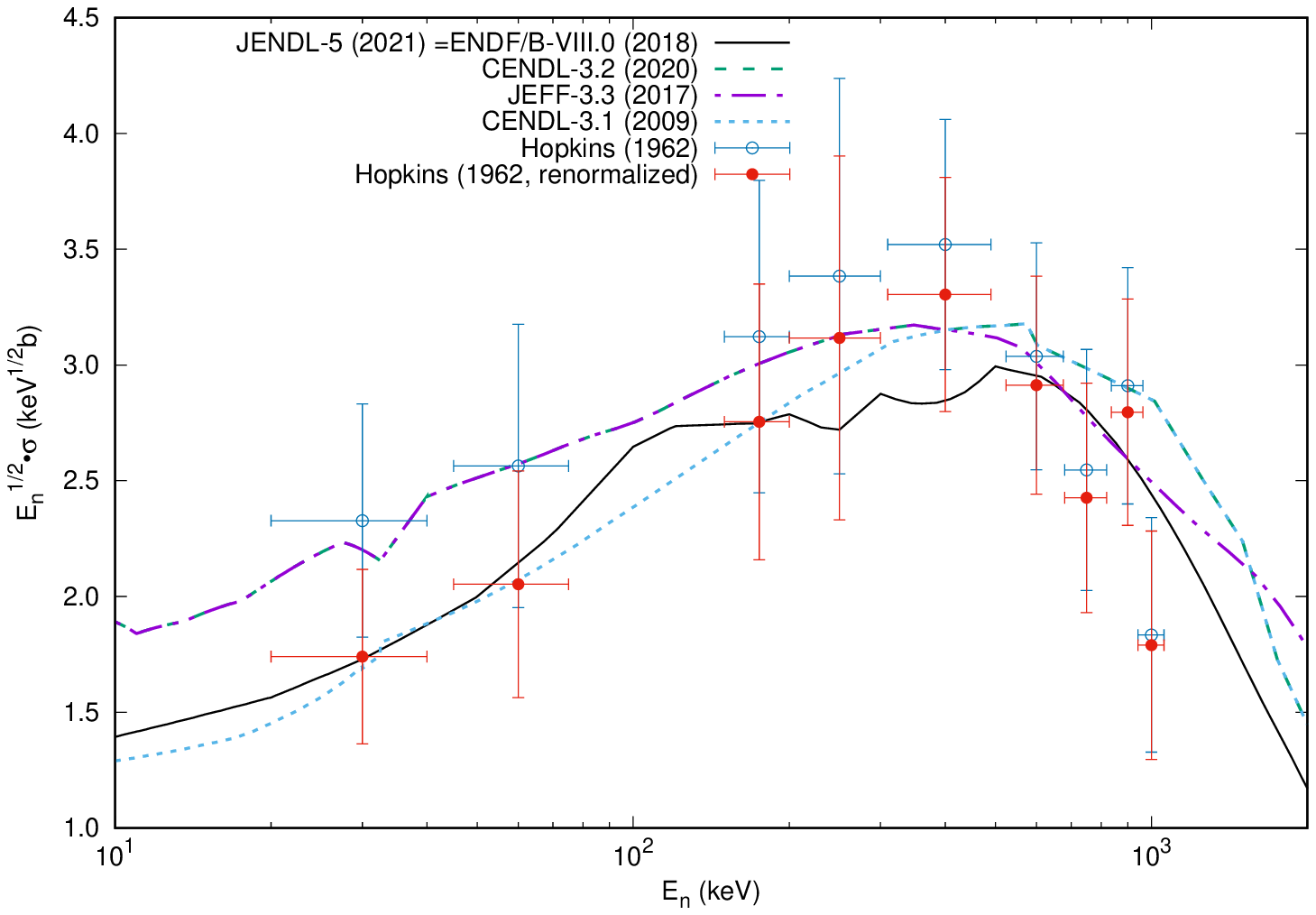}
\caption{
Comparison of $^{233}$U(n,$\gamma$)$^{234}$U cross sections (scaled by the square root of the incident energy) in the latest versions of the major general purpose libraries (JENDL-5, CENDL-3.2, JEFF-3.3) and CENDL-3.1 library with the capture cross section determined by Hopkins et al. (original values published in 1962 and those renormalized with the JENDL-5 $^{233}$U(n,f) cross section).
The ENDF/B-VIII.0 library adopts the JENDL-5 cross section.
The CENDL-3.2 cross section is equal to the JEFF-3.3 cross section (which was originally evaluated for the ENDF/B-VII.0 library) below 400~keV and to the CENDL-3.1 cross section above 400~keV.
}
\label{fig:hopkins}
\end{figure}

Currently three different evaluations of the capture cross section between 3~keV and 1~MeV are seen in the latest versions of the major general purpose libraries:
\begin{itemize}
\item Evaluation adopted in the JENDL-5 library, which was originally obtained with the CCONE code~\cite{Iwamoto2007Development} for the JENDL-4.0 library ~\cite{Shibata2011Japanese} and also adopted in the ENDF/B-VIII.0 library.
\item Evaluation adopted in the JEFF-3.3 library,
which was originally obtained with the GNASH code~\cite{Young1998Comprehensive} for the ENDF/B-VII.0 library~\cite{Chadwick2006ENDF} 
and also adopted in the ENDF/B-VII.1~\cite{Chadwick2011ENDF} and JEFF-3.2~\cite{Plompen2020Joint} libraries as well as the CENDL-3.2 library below 400~keV.
\item Evaluation adopted in the CENDL-3.2 library above 400~keV, which was originally obtained for the CENDL-3.1 library~\cite{Yu1999Evaluation} with the FUNF code~\cite{Zhang2006User}.
\end{itemize}
Figure~\ref{fig:hopkins} shows comparison of these three evaluations with the cross section determined by Hopkins et al.~\cite{Hopkins1962Neutron} which was the only available experimental dataset above 10~keV when these evaluations were performed.
We see in 10~keV region that JEFF-3.3 follows the experimental cross section determined by Hopkins et al., and JENDL-5 follows the same experimental cross section renormalized with the $^{233}$U(n,f) cross section in the JENDL-5 library.
None of the evaluations try to follow the energy dependence of the experimental cross section near 1~MeV.

\section{Present evaluation}
We evaluated the capture cross section from 3~keV to 1~MeV by using the alpha values measured by Hopkins et al. and Leal-Cidoncha et al. by using the least-squares fitting code SOK~\cite{Kawano2000evaluation, Kawano2000Simultaneous}.
We adopted their alpha values in the EXFOR library as inputs to SOK without normalization to the capture cross sections,
and it makes our evaluation procedure free from dependence on choice of the $^{233}$U(n,f) reference cross section seen in Fig.~\ref{fig:hopkins}.

\subsection{Experimental database}
The experimental $^{233}$U alpha value datasets used in the present evaluation are summarized in Table~\ref{tab:explist1}.
Hopkins et al. irradiated a $^{233}$U sample by 30~keV $^7$Li(p,n)$^7$Be neutrons and 60, 175, 250, 400, 600, 750, 900 and 1000~keV $^3$H(p,n)$^3$He neutrons.
They detected $\gamma$-rays from capture and fission events by a cylindrical cadmium-loaded liquid scintillator tank (1~m long and 1~m in diameter).
Fission events were identified by delayed pulses caused by captures of fission neutrons by cadmium after thermalization.
The EXFOR library provides the alpha values and point-wise uncertainties (standard deviations), which are due to statistics in the number of events and due to extrapolation of the pulse-height spectra to zero energy.
The correction factor for the zero energy extrapolation is 1.035$\pm$0.035, and we constructed a covariance matrix of this dataset assuming that its fractional uncertainty ($\sim$3.4\%) is propagated to the point-wise uncertainty of the alpha value as a fully-correlated one.

Leal-Cidoncha et al. irradiated a $^{233}$U sample by spallation neutrons from a tungsten target irradiated by a 800~MeV proton beam.
This measurement used an array of 160 BaF$_2$ crystals (DANCE detector) for prompt gamma detection and another detector consisting of 21 stilbene crystals (NEUANCE detector) for fission neutron detection.
The detector system was located at about 20~m from the spallation target, and the incident energy of each neutron was determined by the time-of-flight technique.
The overall normalization of the measured alpha value was done with the 8.1~eV-14.7~eV capture-to-fission cross section integral ratio in the ENDF/B-VIII.0 library.
The EXFOR library provides the alpha value and point-wise statistical uncertainty (including the uncertainty due to subtraction of the background caused by fission events) as well as the constant uncertainties due to (1) normalization to the ENDF/B-VIII.0 at 8.1~eV-14.7~eV (6.3\%) and (2) normalization in correction of the capture events due to scattered neutrons captured by sample backing and surrounding materials (2\%).
We constructed a covariance matrix of this dataset assuming that these two constant uncertainties as fully-correlated ones.
\begin{table}[hbtp]
\caption{
Summary of the $^{233}$U alpha value datasets included in the experimental database.
``Ver.", ``Lab." and ``Pts." give the date (N2) of the SUBENT record in EXFOR, EXFOR/CINDA abbreviation (see Table 3 of ~\cite{OECD2007CINDA} for its definition) of the institute where the experiment was performed, and number of data points, respectively.
}
\label{tab:explist1}
\begin{tabular}{lllllrccc}
\hline
EXFOR $\#$ &Ver.    &First author   &Year&Lab.   &Pts.&\multicolumn{2}{c}{Energy range (eV)}&Ref.          \\
\hline                                                                                                                        
12331.002  &20231017&J.C.Hopkins    &1962&1USALAS&   9&3.0E+04&1.0E+06&\cite{Hopkins1962Neutron}           \\
14819.002  &20240314&E.Leal-Cidoncha&2023&1USALAS&  55&3.1E+03&2.4E+05&\cite{Leal-Cidoncha2023Measurement}\\
\hline
\end{tabular}
\end{table}

The experimental database of the $^{233,235,238}$U and $^{239,240,241}$Pu fission cross sections and their ratios originally developed for JENDL-5 evaluation~\cite{Otuka2022bEXFOR} is being continuously updated after release of the JENDL-5 library.
Table~\ref{tab:explist2} summarizes the datasets added to the experimental database after JENDL-5 evaluation.
In parallel with these additions,
we also excluded one dataset (one data point) of the $^{235}$U(n,f) cross section measured by Arlt et al.~\cite{Arlt1981Absolute} from the experimental database after the JENDL-5 evaluation as discussed in Ref.~\cite{Otuka2023Simultaneous}.

The present simultaneous fitting was performed for the experimental data points
from 3~keV to 1~MeV for the $^{233}$U(n,$\gamma$)$^{234}$U cross sections,
from 3~keV to 250~MeV for the $^{233,235}$U(n,f) cross sections,
from 7~keV to 250~MeV for the $^{239,241}$Pu(n,f) cross sections,
and from 70~keV to 250~MeV for the $^{238}$U and $^{240}$Pu(n,f) cross sections.
%
\begin{table}[hbtp]
\caption{
Summary of the datasets of $^{233,235,238}$U and $^{239,240,241}$Pu fission cross sections and their ratios
added to the experimental database after JENDL-5 simultaneous evaluation~\cite{Otuka2022bEXFOR}.
See the caption of Table~\ref{tab:explist1} for ``Ver.", ``Lab." and ``Pts.".
}
\label{tab:explist2}
\begin{tabular}{llllllrccc}
\hline
Reaction            &EXFOR $\#$ &Ver.    &First author   &Year&Lab.   &Pts.&\multicolumn{2}{c}{Energy range (eV)}&Ref.         \\
\hline                                                                                                                        
$^{233}$U/$^{235}$U &23654.003.1&20230502&D.Tarr\'{i}o   &2023&2ZZZCER& 110&1.0E+04&2.3E+08&\cite{Tarrio2023Neutron}           \\
$^{235}$U           &23294.004  &20211216&I.Duran        &2019&2ZZZCER&  15&3.4E+03&2.7E+04&\cite{Duran2019High}               \\
$^{238}$U           &51014.002  &20221222&F.Belloni      &2022&2GERPTB&   4&2.5E+06&1.5E+07&\cite{Belloni2022Neutron}          \\
$^{238}$U/$^{235}$U &32886.003.1&20230206&Z.Ren          &2023&3CPRIHP& 135&5.1E+05&1.8E+08&\cite{Ren2023Measurement}          \\
$^{238}$U/$^{235}$U &23657.003.1&20230801&V.Michalopoulou&2023&2ZZZCER& 125&8.0E+05&3.0E+07&\cite{Michalopoulou2023Measurement}\\
$^{238}$U/$^{235}$U &41756.003.1&20231018&A.S.Vorobyev   &2023&4RUSLIN& 176&3.1E+05&2.4E+08&\cite{Vorobyev2023Measurement}     \\
$^{239}$Pu/$^{235}$U&14721.002  &20211220&L.Snyder       &2021&1USALAS& 119&1.1E+05&9.7E+07&\cite{Snyder2021Measurement}       \\
$^{240}$Pu          &23653.002.1&20221126&F.Belloni      &2022&2GERPTB&   2&2.5E+06&1.5E+07&\cite{Belloni2022Neutron}          \\
\hline
\end{tabular}
\end{table}

\subsection{Evaluation method}
Details of the current evaluation procedure are summarized in our previous publications~\cite{Otuka2022aEXFOR, Kawano2000evaluation, Kawano2000Simultaneous, Otuka2023Simultaneous},
and we provide below a short summary of the formalism.
We model energy dependence of the logarithmic of an experimental cross section $\Sigma_\mathrm{exp}(E_n)=\ln\left[\sigma_\mathrm{exp}(E_n)\right]$ by linear combinations of the evaluated cross sections at the nearest two energy grids $E_{j-1}$ and $E_j$:
\begin{equation}
\Sigma_\mathrm{exp}(E_n)=\Delta_{j-1}(E_n)\Sigma_\mathrm{eva}(E_{j-1})+\Delta_j(E_n)\Sigma_\mathrm{eva}(E_j)+\delta
\label{eqn:lincom1}
\end{equation}
($E_{j-1}\le E_n<E_j$),
where $\delta$ is the residual of fitting and $\Delta_j(E_n)$ is the roof function~\cite{Schmittroth1980Finite} defined by
\begin{equation}
\Delta_j(E_n)=
\left\{
\renewcommand{\arraystretch}{1.2}
\begin{array}{ll}
\frac{E_n-E_{j-1}}{E_j-E_{j-1}}& (E_{j-1} \le E_n \ < E_j)     \\
1                               & ( E_n = E_j)                \\
\frac{E_{j+1}-E_n}{E_{j+1}-E_j}& (E_j     <  E_n \ < E_{j+1})\\
0                               & \textrm{otherwise}          \\
\end{array}
\renewcommand{\arraystretch}{1}
\right.
.
\end{equation}
When $E_n$ is the middle of $E_{j-1}$ and $E_j$, Eq.~(\ref{eqn:lincom1}) is simplified to
\begin{equation}
\Sigma_\mathrm{exp}(E_n)=0.5\,\Sigma_\mathrm{eva}(E_{j-1})+0.5\,\Sigma_\mathrm{eva}(E_j)+\delta.
\end{equation}
If there is a ratio of the cross sections for the reaction A to B measured at $E_n$,
its logarithmic $\Sigma^{A/B}_\mathrm{exp}(E_n)$ is related with the evaluated cross sections for the reaction A at $E_{j-1}$ and $E_j$ and those for the reaction B at $E_{k-1}$ and $E_k$ by
\begin{equation}
\Sigma^{A/B}_\mathrm{exp}(E_n)=
\left[\Delta_{j-1}(E_n)\Sigma^A_\mathrm{eva}(E_{j-1})+\Delta_j(E_n)\Sigma^A_\mathrm{eva}(E_j)\right] -
\left[\Delta_{k-1}(E_n)\Sigma^B_\mathrm{eva}(E_{k-1})+\Delta_k(E_n)\Sigma^B_\mathrm{eva}(E_k)\right]+\delta.
\label{eqn:lincom2}
\end{equation}
These equations imply that the logarithmics of the experimental cross sections or cross section ratios ($m$ experimental data points) can be expressed by the logarithmics of the evaluated cross sections ($n$ evaluated cross sections) by 
\begin{equation}
\vec{\Sigma}_\mathrm{exp}=C \vec{\Sigma}_\mathrm{eva}+\vec{\delta},
\label{eqn:glsq}
\end{equation}
where $\vec{\Sigma}_\mathrm{exp}$ and $\vec{\delta}$ are $m$-dimensional vectors, $\vec{\Sigma}_\mathrm{eva}$ is a $n$-dimensional vector, and $C$ is a $n\times m$ matrix.
The SOK code does not solve Eq.~(\ref{eqn:glsq}) directly,
but obtains $\vec{\Sigma}_\mathrm{eva}$=$\vec{\Sigma}_\mathrm{eva}^N$ and its covariance $V_\mathrm{eva}$=$V_\mathrm{eva}^N$ by updating their initial guess (prior value) $\vec{\Sigma}_\mathrm{eva}^0$ and $V_\mathrm{eva}^0$ to $\vec{\Sigma}_\mathrm{eva}^N$ and $V_\mathrm{eva}^N$ (posterior value) following the generalized least-squares approach~\cite{Mannhart2013Small,Hirtz2024Parameter}:
\begin{eqnarray}
\vec{\Sigma}_\mathrm{eva}^i&=&\vec{\Sigma}_\mathrm{eva}^{i-1}+V_\mathrm{eva}^{i-1}\,C^{i\mathsf{T}}(U^i+V_\mathrm{exp}^i)^{-1}(\vec{\Sigma}_\mathrm{exp}^i-C^i\vec{\Sigma}_\mathrm{eva}^{i-1}),\\
V_\mathrm{eva}^i           &=&V_\mathrm{eva}^{i-1}-V_\mathrm{eva}^{i-1}\,C^{i\mathsf{T}}(U^i+V_\mathrm{exp}^i)^{-1}C^iV_\mathrm{eva}^{i-1}
\end{eqnarray}
where 
$
\vec{\Sigma}_\mathrm{exp}^\mathsf{T}=
\left(
 \vec{\Sigma}_\mathrm{exp}^{1\mathsf{T}},\,
 \vec{\Sigma}_\mathrm{exp}^{2\mathsf{T}},\,
 \cdots,
 \vec{\Sigma}_\mathrm{exp}^{N\mathsf{T}}
\right)
$,
$V_\mathrm{exp}=\mathrm{diag}(V_\mathrm{exp}^1, V_\mathrm{exp}^2, \cdots, V_\mathrm{exp}^N$),
$C^i=\mathrm{diag}(C^1, C^2, \cdots, C^N)$, and
$U^i=C^iV_\mathrm{eva}^{i-1}C^{i\mathsf{T}}$.
The $^{233}$U(n,$\gamma$)$^{234}$U cross section in the JENDL-5 library and the $^{233,235,238}$U and $^{239,240,241}$Pu fission cross sections from the JENDL-5 simultaneous evaluation were adopted as $\vec{\Sigma}_\mathrm{eva}^0$ with 50\% uncorrelated uncertainty for $V_\mathrm{eva}^0$\footnote{
The JENDL-5 library covariance could be an alternative of the prior input. If one chooses this option, the experimental covariances used in JENDL-5 evaluation cannot be used again to update the JENDL-5 cross section.
}.

This formalism allows inclusion of experimental cross sections and cross section ratios for many target nuclides and reaction channels.
For example, the $^{233}$U(n,$\gamma$)$^{234}$ cross section can be evaluated without any experimental dataset of the absolute $^{233}$U(n,$\gamma$)$^{234}$ cross section 
if we have enough experimental datasets of the $^{233}$U(n,$\gamma$) /$^{233}$U(n,f) and $^{233}$U(n,f) /$^{235}$U(n,f) cross section ratios as well as the absolute $^{235}$U(n,f) cross section.

\subsection{Evaluation result}
Fitting to the roof function~\cite{Schmittroth1980Finite} was done for 8178 experimental data points and 562 fitting parameters (559 parameters of evaluated cross sections and three parameters of the overall normalization factors for the three experimental datasets) with the reduced chi-square of 5.88.
Among the 8178 experimental data points, (1) 64 data points are summarized in Table 1, (2) 686 data points are summarized in Table 2, (3) 7378 data points are taken from the measurements summarized in Tables 3 to 15 of~\cite{Otuka2022bEXFOR} and used in the JENDL-5 simultaneous evaluation (excluding one point measured by Arlt et al.~\cite{Arlt1981Absolute}), and (4) 50 data points are from the measurements in Tables 3 to 15 of~\cite{Otuka2022bEXFOR} but below the lower energy boundary of the JENDL-5 simultaneous evaluation. The 559 incident energies of the evaluated cross sections were chosen to reproduce the energy dependences of the experimental data points reasonably.

\begin{figure}[hbtp]
\includegraphics[angle=-90,width=\linewidth]{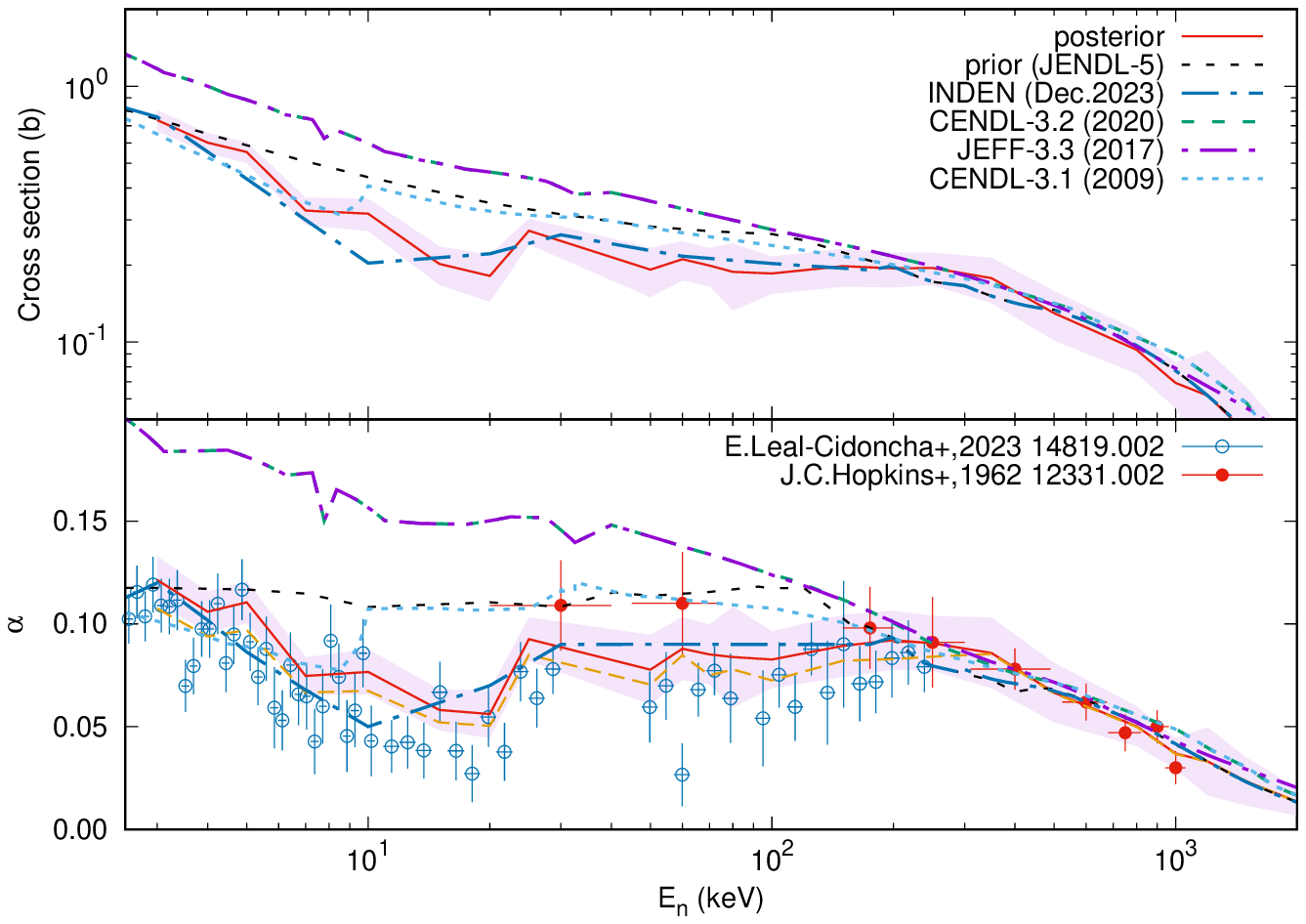}
\caption{
Comparison of the newly evaluated $^{233}$U(n,$\gamma$)$^{234}$U cross section (top) and alpha value (bottom) with those measured by Hopkins et al. and Leal-Cidoncha et al. and those in evaluated data libraries.
The pink shaded area displays the internal uncertainty in the present evaluation with the roof function.
The dashed line close to the lower boundary of the pink shaded area in the bottom panel shows the posterior value obtained by fitting without correlation between data points measured by Leal-Cidoncha et al.
}
\label{fig:evasigcap}
\end{figure}

Figure~\ref{fig:evasigcap} shows the capture cross section and alpha value from the current evaluation\footnote{
A table of the newly evaluated cross sections is available as a Supplemental Material.
}.
The uncertainty in the present evaluation with the roof function is displayed by the pink shaded area, which is the internal uncertainty and must be enlarged by $\sqrt{5.88}\sim 2.42$ to achieve the reduced chi-square equal to 1.
The uncertainty in the evaluated cross section is too high to meet the proposed target accuracies mentioned in the introduction (5\% or 9\%).
As discussed in Ref.~\cite{Leal-Cidoncha2023Measurement},
the newly measured alpha value is lower than those measured by Hopkins et al. below 100~keV,
while they are consistent above 200~keV.
Unfortunately,
the new measurement does not provide alpha values between 30 and 48~keV due to reduction of the neutron flux.
It makes the influence of the alpha values measured by Hopkins et al. stronger and also makes the uncertainty of the current evaluation large in this energy region.

Figure~\ref{fig:evasigcap} shows that the newly evaluated capture cross section and alpha value are systematically lower than their values in the JENDL-5 library.
This figure also shows that the present evaluation is close to the INDEN evaluation~\cite{Pigni2023INDEN}, which takes into account the measurement by Leal-Cidoncha et al.

The posterior alpha value below 20~keV seems higher than the majority of the data points measured by Leal-Cidoncha et al.
To understand this situation,
we performed additional fitting without the off-diagonal elements of the correlation coefficients for the dataset measured by Leal-Cidoncha et al.,
and obtained a lower posterior alpha value close to the lower boundary of the uncertainty band of the alpha value from the original fitting.
Namely,
elimination of the off-diagonal elements in the new experimental dataset systematically reduces the posterior alpha value.
The measured data points with higher alpha values tend to have lower uncertainties and are more preferred by the least-squares fitting.
Such data points may have influence to normalization of the posterior if all data points of the dataset are correlated.
This would explain systematic reduction of the alpha value by elimination of the correlation.

\begin{figure}[hbtp]
\includegraphics[angle=-90,width=\linewidth]{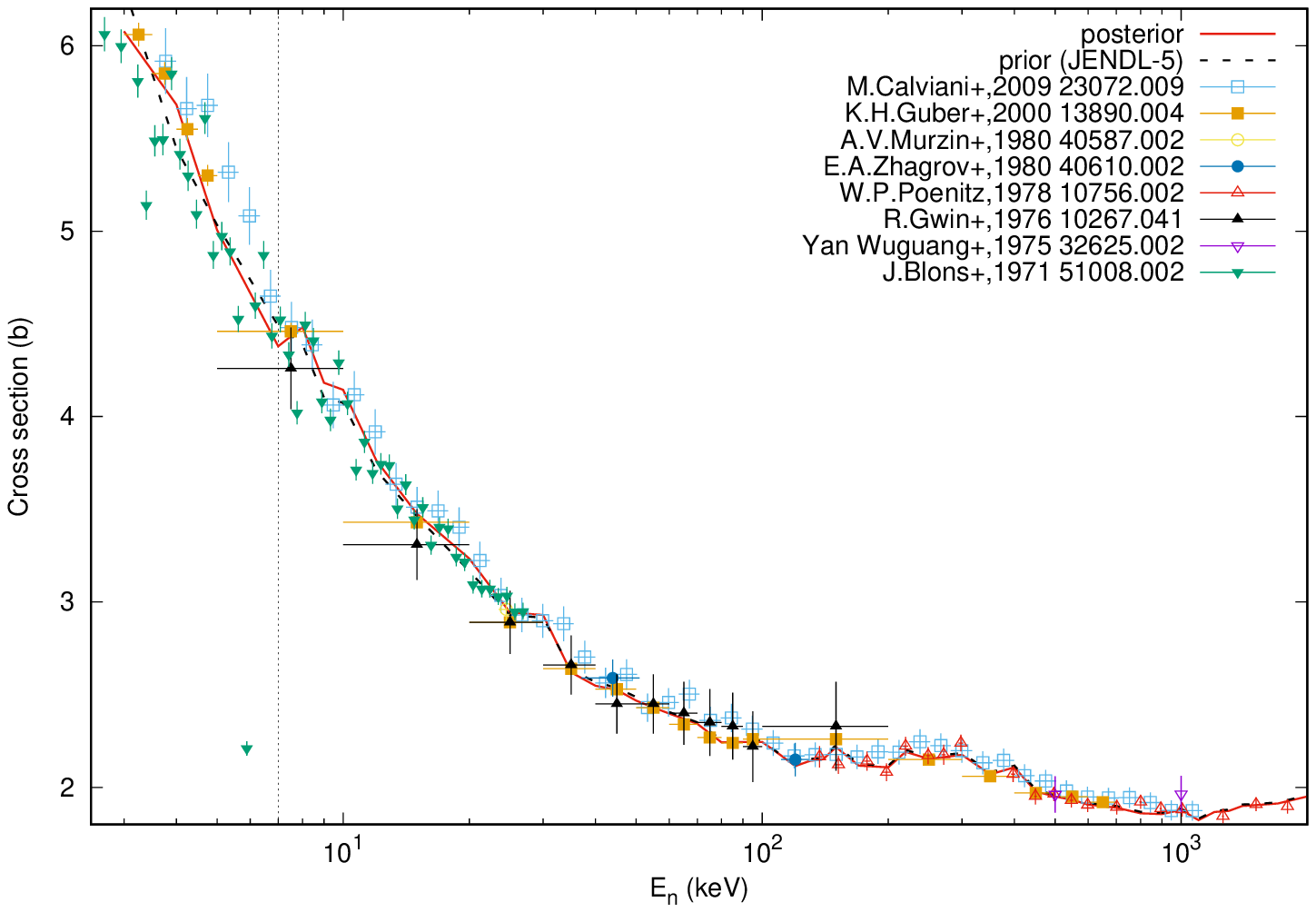}
\caption{
Comparison between $^{233}$U(n,f) cross sections before and after update with the $^{233}$U alpha values measured by Hopkins et al. and Leal-Cidoncha et al. as well as the experimental datasets listed in Table~\ref{tab:explist2}.
}
\label{fig:evasigfis}
\end{figure}

Figure~\ref{fig:evasigfis} shows influence of inclusion of the alpha values measured by Hopkins et al. and Leal-Cidoncha et al. as well as the datasets added after JENDL-5 evaluation (Table~\ref{tab:explist2}) to the posterior $^{233}$U(n,f) cross section.
The experimental fission cross sections below 7~keV were not used in fitting for the JENDL-5 library but used in the present evaluation.
We see that inclusion of the experimental alpha values in the fitting procedure does not change the fission cross section above 25~keV and introduces very minor between 7~keV and 25~keV, which may be partly due to addition of the fission datasets listed in Table~\ref{tab:explist2} (e.g., \cite{Tarrio2023Neutron}) rather than addition of the alpha values.
Below 7~keV, the JENDL-5 cross section adopts the JENDL-4.0 cross section, which was analysis of the experimental data measured after 1960 by the GMA code~\cite{Poenitz1997Simultaneous}.
This figure shows that fitting with the experimental alpha values in this region prefers the fission cross section measured by Guber et al.~\cite{Guber2000New} rather than Blons et al.~\cite{Blons1971MeasurementAnalysis}.

\section{Benchmark}
Benchmark calculations were performed by the continuous-energy Monte Carlo neutron transport code MVP version 3 (MVP3)~\cite{Nagaya2017MVP} with the MVP input files for the 166 criticality experiments~\cite{Kuwagai2017Integral,JENDL2023Compilation} in the ICSBEP handbook~\cite{OECD2014International}.
These experiments for uranium-233 systems were chosen for the JENDL-5 benchmark test~\cite{Tada2024JENDL-5}.
Table~\ref{tab:ICSBEP} summarizes the number of experiments adopted in the current benchmark (See Tables \ref{tab:UCT} to \ref{tab:UST} for more detailed listing of all cases).
For $^{233}$U, the JENDL-5 file was updated with the newly evaluated capture cross section between 3~keV and 1~MeV by using DeCE~\cite{Kawano2019Dece} and processed by LICEM~\cite{Mori1996Neutron} for the MVP calculation.
The JENDL-5 file compiles the capture cross section in File 3 with LSSF=1 flag in the ENDF-6 format, namely File 3 includes the entire capture cross section in the unresolved resonance region.
For the other nuclides, the original JENDL-5 files processed for its benchmarking~\cite{Tada2024JENDL-5} were used.
%
\begin{table}[hbtp]
\caption{
Number of benchmark experiments in the ICSBEP handbook and adopted in the present benchmark.
}
\label{tab:ICSBEP}
\begin{tabular}{lrlrlrlrlrlr}
\hline
UCT001 &  8 & UMF003 &  2 & USI001 & 30 & UST002 & 17 & UST008 &  1 & UST015 &  26 \\
UCT004 &  1 & UMF004 &  2 & USM001 &  3 & UST003 & 10 & UST009 &  4 & UST017 &   7 \\
UMF001 &  1 & UMF005 &  2 & USM002 &  5 & UST004 &  8 & UST012 &  8 &        &     \\
UMF002 &  2 & UMF006 &  1 & UST001 &  5 & UST005 &  2 & UST013 & 21 & Total  & 166 \\
\hline
\end{tabular}
\end{table}

\begin{figure}[hbtp]
\includegraphics[angle=-90,width=\linewidth]{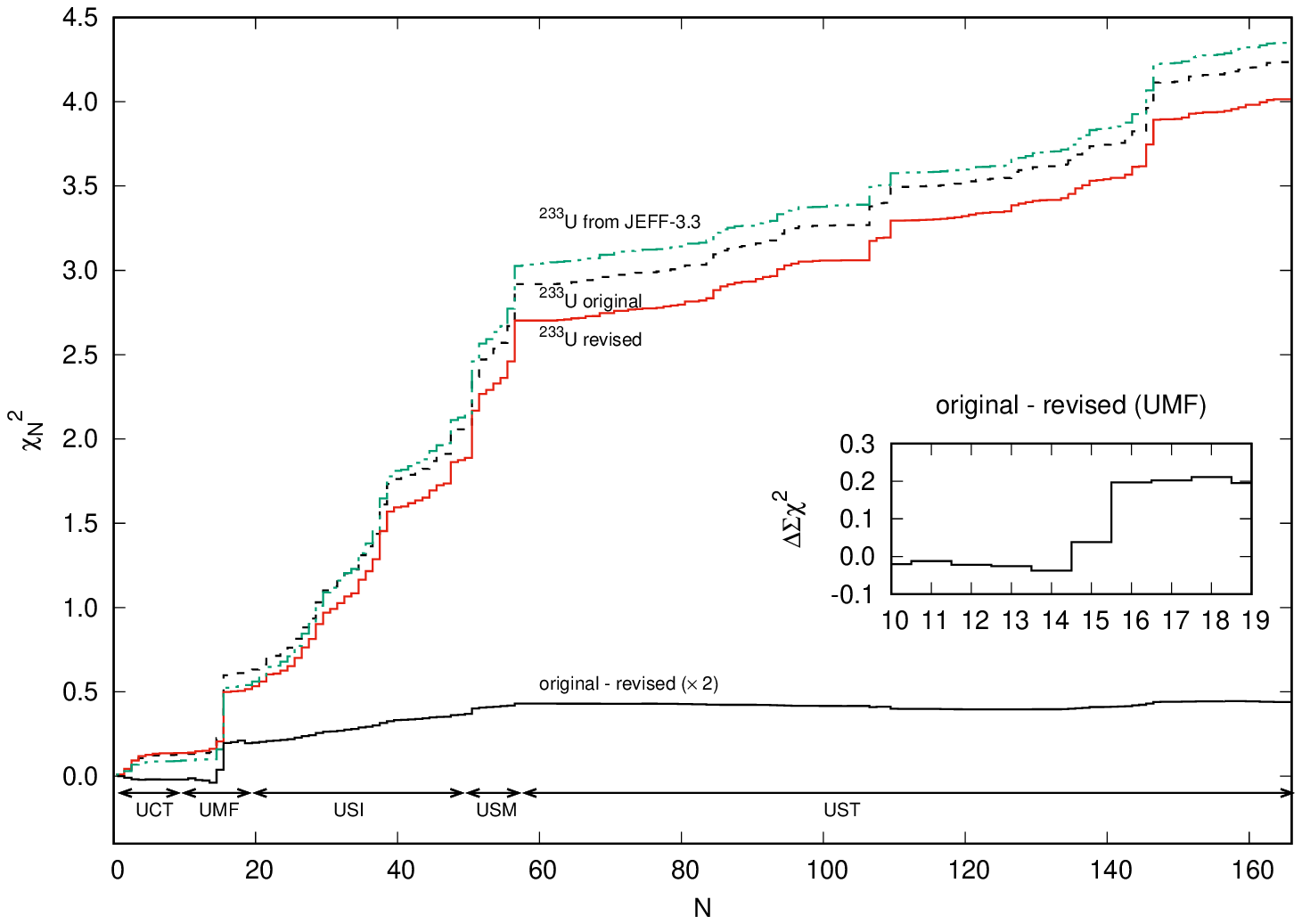}
\caption{
Cumulative reduced chi-square values for the benchmark calculations by MVP3 with the JENDL-5 library for the 166 cases of the $^{233}$U systems~\cite{Kuwagai2017Integral,JENDL2023Compilation} chosen from the ICSBEP handbook.
Three curves annotated by $^{233}$U original, $^{233}$U revised, and $^{233}$U from JEFF-3.3) are for the benchmark calculations with the original JENDL-5 $^{233}$U file, revised JENDL-5 $^{233}$U file and JEFF-3 $^{233}$U file, respectively.
The curve for benchmark calculation with the ENDF/B-VIII.0 $^{233}$U file is similar to the curve with the original JENDL-5 $^{233}$U file,
and omitted for clarity.
}
\label{fig:MVP}
\end{figure}

To see an overall agreement between the calculated value $C_i$ and experimental value $E_i$,
we estimated the reduced chi-square defined by
\begin{equation}
\chi_N^2=\frac{1}{N-1}\sum_{i=1}^{N} \frac{\left(C_i/E_i-1\right)^2}{(\delta C_i)^2+(\delta E_i)^2},
\end{equation}
where $N$=166 and $\delta C_i$ and $\delta E_i$ are the fractional uncertainties in the calculated and experimental values for the experiment $i$, respectively.
We found that $\chi^2$=4.26 for the original JENDL-5 library,
and it is slightly reduced to 4.04 by adoption of the newly evaluated $^{233}$U capture cross section.
Figure~\ref{fig:MVP} is the cumulative reduced chi-square values for the benchmark calculations with the original and revised JENDL-5 $^{233}$U files.
The largest contribution to the improvement in the $\chi^2$ values is seen for two metal fast systems,  UMF004-001 (Case\# 15) and UMF004-002 (Case\# 16).
Systematic improvements are also seen in $\chi^2$ values of solution intermediate systems (USI) and solution mixed systems (USM).
To see the performance of the $^{233}$U files of other major general purpose libraries,
we additionally performed the same benchmark calculation by replacing the JENDL-5 $^{233}$U file with those in the ENDF/B-VIII.0 and JEFF-3.3,
and obtained $\chi^2$=4.27 and 4.38, respectively.

\section{Summary}
We evaluated the $^{233}$U neutron capture cross sections between 3~keV and 1~MeV considering the new alpha-value measurement performed at the Los Alamos National Laboratory LANCE facility within the simultaneous least-squares analysis framework including the $^{233,235,238}$U and $^{239,240,241}$Pu fission cross sections.
The obtained capture cross section is systematically lower than those in the latest versions of the major general purpose nuclear data libraries,
and the reduction from the JENDL-5 library is close to 50\% around 20~keV.

We benchmarked the newly evaluated cross section by performing Monte Carlo neutron transport calculation with the JENDL-5 library for 166 criticality experiments chosen from the ICSBEP handbook,
and found slight reduction in the chi-square value by adoption of the newly evaluated capture cross section.

The experimental information of the $^{233}$U neutron capture cross section in the keV region is still insufficient to meet the target accuracy for reactor applications,
and additional measurements are desired.

\section*{Acknowledgement}
We would like to thank Yasunobu Nagaya (JAEA) for processing of the revised JENDL-5 ENDF file for MVP calculation.
The EXFOR entry compiling the newly published alpha values measured at LANSCE (EXFOR \#14819) was prepared by Stanislav Hlav\'{a}\v{c} (Institute of Physics, Slovak Academy of Sciences),
who contributed to the EXFOR library from 2007 to 2023 as an EXFOR compiler of the National Nuclear Data Center (NNDC).
One of us (NO) is also grateful to Wu Haicheng (CIAE) for discussion on the status of the $^{233}$U neutron capture cross section data
and to Esther Leal-Cidoncha (LANL) for additional explanation on the partial uncertainties compiled in the EXFOR entry.

\appendix
\section{Full list of cases in benchmark calculation}
The ICSBEP $^{233}$U cases used in the benchmark calculations performed in the present work are summarized in Tables \ref{tab:UCT} to \ref{tab:UST} with their case numbers used in Fig.~\ref{fig:MVP}.

\begin{table}[hbtp]
\caption{
Benchmark cases for $^{233}$U compound thermal systems (UCT)
}
\label{tab:UCT}
\begin{tabular}{rlrlrl}
\hline
 1 &UCT001-001 & 4 &UCT001-004 & 7 &UCT001-007 \\
 2 &UCT001-002 & 5 &UCT001-005 & 8 &UCT001-008 \\
 3 &UCT001-003 & 6 &UCT001-006 & 9 &UCT004     \\
\hline
\end{tabular}
\end{table}

\begin{table}[hbtp]
\caption{
Benchmark cases for $^{233}$U metal fast systems (UMF)
}
\label{tab:UMF}
\begin{tabular}{rlrlrlrl}
\hline
 10 &UMF001-001 & 13 &UMF003-001 & 16 &UMF004-002 & 19 &UMF006-001 \\
 11 &UMF002-001 & 14 &UMF003-002 & 17 &UMF005-001 &    &           \\
 12 &UMF002-002 & 15 &UMF004-001 & 18 &UMF005-002 &    &           \\
\hline
\end{tabular}
\end{table}

\begin{table}[hbtp]
\caption{
Benchmark cases for $^{233}$U solution intermediate systems (USI)
}
\label{tab:USI}
\begin{tabular}{rlrlrlrlrlrl}
\hline
 20 &USI001-001& 28 &USI001-009 & 36 &USI001-019 & 44 &USI001-027 \\
 21 &USI001-002& 29 &USI001-010 & 37 &USI001-020 & 45 &USI001-028 \\
 22 &USI001-003& 30 &USI001-011 & 38 &USI001-021 & 46 &USI001-029 \\
 23 &USI001-004& 31 &USI001-012 & 39 &USI001-022 & 47 &USI001-031 \\
 24 &USI001-005& 32 &USI001-013 & 40 &USI001-023 & 48 &USI001-032 \\
 25 &USI001-006& 33 &USI001-015 & 41 &USI001-024 & 49 &USI001-033 \\
 26 &USI001-007& 34 &USI001-017 & 42 &USI001-025 &    &           \\
 27 &USI001-008& 35 &USI001-018 & 43 &USI001-026 &    &           \\
\hline
\end{tabular}
\end{table}

\begin{table}[hbtp]
\caption{
Benchmark cases for $^{233}$U solution mixed systems (USM)
}
\label{tab:USM}
\begin{tabular}{rlrlrlrlrlrl}
\hline
 50 &USM001-014& 52 &USM001-030& 54 &USM002-005 & 56 &USM002-008 \\
 51 &USM001-016& 53 &USM002-003& 55 &USM002-006 & 57 &USM002-009 \\
\hline
\end{tabular}
\end{table}

\begin{table}[hbtp]
\caption{
Benchmark cases for $^{233}$U solution thermal systems (UST)
}
\label{tab:UST}
\begin{tabular}{rlrlrlrlrlrl}
\hline
 58 &UST001-001 & 86 &UST003-007 &114 &UST013-002 &142 &UST015-014 \\
 59 &UST001-002 & 87 &UST003-008 &115 &UST013-003 &143 &UST015-015 \\
 60 &UST001-003 & 88 &UST003-009 &116 &UST013-004 &144 &UST015-016 \\
 61 &UST001-004 & 89 &UST003-010 &117 &UST013-005 &145 &UST015-017 \\
 62 &UST001-005 & 90 &UST004-001 &118 &UST013-006 &146 &UST015-018 \\
 63 &UST002-001 & 91 &UST004-002 &119 &UST013-007 &147 &UST015-019 \\
 64 &UST002-002 & 92 &UST004-003 &120 &UST013-008 &148 &UST015-020 \\
 65 &UST002-003 & 93 &UST004-004 &121 &UST013-009 &149 &UST015-021 \\
 66 &UST002-004 & 94 &UST004-005 &122 &UST013-010 &150 &UST015-022 \\
 67 &UST002-005 & 95 &UST004-006 &123 &UST013-011 &151 &UST015-023 \\
 68 &UST002-006 & 96 &UST004-007 &124 &UST013-012 &152 &UST015-024 \\
 69 &UST002-007 & 97 &UST004-008 &125 &UST013-013 &153 &UST015-025 \\
 70 &UST002-008 & 98 &UST005-001 &126 &UST013-014 &154 &UST015-026 \\
 71 &UST002-009 & 99 &UST005-002 &127 &UST013-015 &155 &UST015-027 \\
 72 &UST002-010 &100 &UST008-001 &128 &UST013-016 &156 &UST015-028 \\
 73 &UST002-011 &101 &UST009-001 &129 &UST013-017 &157 &UST015-029 \\
 74 &UST002-012 &102 &UST009-002 &130 &UST013-018 &158 &UST015-030 \\
 75 &UST002-013 &103 &UST009-003 &131 &UST013-019 &159 &UST015-031 \\
 76 &UST002-014 &104 &UST009-004 &132 &UST013-020 &160 &UST017-001 \\
 77 &UST002-015 &105 &UST012-001 &133 &UST013-021 &161 &UST017-002 \\
 78 &UST002-016 &106 &UST012-002 &134 &UST015-001 &162 &UST017-003 \\
 79 &UST002-017 &107 &UST012-003 &135 &UST015-002 &163 &UST017-004 \\
 80 &UST003-001 &108 &UST012-004 &136 &UST015-004 &164 &UST017-005 \\
 81 &UST003-002 &109 &UST012-005 &137 &UST015-007 &165 &UST017-006 \\
 82 &UST003-003 &110 &UST012-006 &138 &UST015-010 &166 &UST017-007 \\
 83 &UST003-004 &111 &UST012-007 &139 &UST015-011 &    &           \\
 84 &UST003-005 &112 &UST012-008 &140 &UST015-012 &    &           \\
 85 &UST003-006 &113 &UST013-001 &141 &UST015-013 &    &           \\
\hline
\end{tabular}
\end{table}
\clearpage
\bibliography{u233-ane}

\begin{thebibliography}{49}
\expandafter\ifx\csname natexlab\endcsname\relax\def\natexlab#1{#1}\fi
\providecommand{\url}[1]{\texttt{#1}}
\providecommand{\href}[2]{#2}
\providecommand{\path}[1]{#1}
\providecommand{\DOIprefix}{doi:}
\providecommand{\ArXivprefix}{arXiv:}
\providecommand{\URLprefix}{URL: }
\providecommand{\Pubmedprefix}{pmid:}
\providecommand{\doi}[1]{\href{http://dx.doi.org/#1}{\path{#1}}}
\providecommand{\Pubmed}[1]{\href{pmid:#1}{\path{#1}}}
\providecommand{\bibinfo}[2]{#2}
\ifx\xfnm\relax \def\xfnm[#1]{\unskip,\space#1}\fi
\bibitem[{{International Atomic Energy Agency}(2005)}]{IAEA2005Thorium}
\bibinfo{author}{{International Atomic Energy Agency}}, \bibinfo{title}{Thorium
  fuel cycle --- {P}otential benefits and challenges}, \bibinfo{type}{Technical
  Report} \bibinfo{number}{IAEA-TECDOC-1450}, International Atomic Energy
  Agency, \bibinfo{year}{2005}.
\bibitem[{Pronyaev(1999)}]{Pronyaev1999Summary}
\bibinfo{author}{V.~G. Pronyaev}, \bibinfo{title}{Summary report of the
  {C}onsultants' {M}eeting on {A}ssessment of {N}uclear {D}ata {N}eeds for
  {T}horium and {O}ther {A}dvanced {C}ycles}, \bibinfo{type}{Technical Report}
  \bibinfo{number}{INDC(NDS)-408}, International Atomic Energy Agency,
  \bibinfo{year}{1999}.
\bibitem[{{International Atomic Energy Agency}(2010)}]{IAEA2010Evaluated}
\bibinfo{author}{{International Atomic Energy Agency}},
  \bibinfo{title}{Evaluated nuclear data for nuclides within the
  thorium-uranium fuel cycle}, \bibinfo{type}{Technical Report}
  \bibinfo{number}{STI/PUB/1435}, International Atomic Energy Agency,
  \bibinfo{year}{2010}.
\bibitem[{Dupont et~al.(2020)Dupont, Bossant, Capote, Carlson, Danon, Fleming,
  Ge, Harada, Iwamoto, Iwamoto, Kimura, Koning, Massimi, Negret, Noguere,
  Plompen, Pronyaev, Rimpault, Simakov, Stankovskiy, Sun, Trkov, Wu, and
  Yokoyama}]{Dupont2020HPRL}
\bibinfo{author}{E.~Dupont}, \bibinfo{author}{M.~Bossant},
  \bibinfo{author}{R.~Capote}, \bibinfo{author}{A.~D. Carlson},
  \bibinfo{author}{Y.~Danon}, \bibinfo{author}{M.~Fleming},
  \bibinfo{author}{Z.~Ge}, \bibinfo{author}{H.~Harada},
  \bibinfo{author}{O.~Iwamoto}, \bibinfo{author}{N.~Iwamoto},
  \bibinfo{author}{A.~Kimura}, \bibinfo{author}{A.~J. Koning},
  \bibinfo{author}{C.~Massimi}, \bibinfo{author}{A.~Negret},
  \bibinfo{author}{G.~Noguere}, \bibinfo{author}{A.~Plompen},
  \bibinfo{author}{V.~Pronyaev}, \bibinfo{author}{G.~Rimpault},
  \bibinfo{author}{S.~Simakov}, \bibinfo{author}{A.~Stankovskiy},
  \bibinfo{author}{W.~Sun}, \bibinfo{author}{A.~Trkov},
  \bibinfo{author}{H.~Wu}, \bibinfo{author}{K.~Yokoyama},
\newblock \bibinfo{title}{{HPRL} -- {I}nternational cooperation to identify and
  monitor priority nuclear data needs for nuclear applications},
\newblock \bibinfo{journal}{EPJ Web of Conferences} \bibinfo{volume}{239}
  (\bibinfo{year}{2020}) \bibinfo{pages}{15005}.
  \DOIprefix\doi{10.1051/epjconf/202023915005}.
\bibitem[{Ge et~al.(2020)Ge, Xu, Wu, Zhang, Chen, Jin, Shu, Chen, Tao, Tian,
  Liu, Qian, Wang, Zhang, Liu, and Huang}]{Ge2020CENDL}
\bibinfo{author}{Z.~Ge}, \bibinfo{author}{R.~Xu}, \bibinfo{author}{H.~Wu},
  \bibinfo{author}{Y.~Zhang}, \bibinfo{author}{G.~Chen},
  \bibinfo{author}{Y.~Jin}, \bibinfo{author}{N.~Shu},
  \bibinfo{author}{Y.~Chen}, \bibinfo{author}{X.~Tao},
  \bibinfo{author}{Y.~Tian}, \bibinfo{author}{P.~Liu},
  \bibinfo{author}{J.~Qian}, \bibinfo{author}{J.~Wang},
  \bibinfo{author}{H.~Zhang}, \bibinfo{author}{L.~Liu},
  \bibinfo{author}{X.~Huang},
\newblock \bibinfo{title}{{CENDL-3.2}: The new version of {C}hinese general
  purpose evaluated nuclear data library},
\newblock \bibinfo{journal}{EPJ Web of Conferences} \bibinfo{volume}{239}
  (\bibinfo{year}{2020}) \bibinfo{pages}{09001}.
  \DOIprefix\doi{10.1051/epjconf/202023909001}.
\bibitem[{Brown et~al.(2018)Brown, Chadwick, Capote, Kahler, Trkov, Herman,
  Sonzogni, Danon, Carlson, Dunn, Smith, Hale, Arbanas, Arcilla, Bates, Beck,
  Becker, Brown, Casperson, Conlin, Cullen, Descalle, Firestone, Gaines, Guber,
  Hawari, Holmes, Johnson, Kawano, Kiedrowski, Koning, Kopecky, Leal, Lestone,
  Lubitz, M{\'a}rquez~Dami{\'a}n, Mattoon, McCutchan, Mughabghab, Navratil,
  Neudecker, Nobre, Noguere, Paris, Pigni, Plompen, Pritychenko, Pronyaev,
  Roubtsov, Rochman, Romano, Schillebeeckx, Simakov, Sin, Sirakov, Sleaford,
  Sobes, Soukhovitskii, Stetcu, Talou, Thompson, van~der Marck,
  Welser-Sherrill, Wiarda, White, Wormald, Wright, Zerkle, \v{Z}erovnik, and
  Zhu}]{Brown2018ENDF}
\bibinfo{author}{D.~A. Brown}, \bibinfo{author}{M.~B. Chadwick},
  \bibinfo{author}{R.~Capote}, \bibinfo{author}{A.~C. Kahler},
  \bibinfo{author}{A.~Trkov}, \bibinfo{author}{M.~W. Herman},
  \bibinfo{author}{A.~A. Sonzogni}, \bibinfo{author}{Y.~Danon},
  \bibinfo{author}{A.~D. Carlson}, \bibinfo{author}{M.~Dunn},
  \bibinfo{author}{D.~L. Smith}, \bibinfo{author}{G.~M. Hale},
  \bibinfo{author}{G.~Arbanas}, \bibinfo{author}{R.~Arcilla},
  \bibinfo{author}{C.~R. Bates}, \bibinfo{author}{B.~Beck},
  \bibinfo{author}{B.~Becker}, \bibinfo{author}{F.~Brown},
  \bibinfo{author}{R.~J. Casperson}, \bibinfo{author}{J.~Conlin},
  \bibinfo{author}{D.~E. Cullen}, \bibinfo{author}{M.-A. Descalle},
  \bibinfo{author}{R.~Firestone}, \bibinfo{author}{T.~Gaines},
  \bibinfo{author}{K.~H. Guber}, \bibinfo{author}{A.~I. Hawari},
  \bibinfo{author}{J.~Holmes}, \bibinfo{author}{T.~D. Johnson},
  \bibinfo{author}{T.~Kawano}, \bibinfo{author}{B.~C. Kiedrowski},
  \bibinfo{author}{A.~J. Koning}, \bibinfo{author}{S.~Kopecky},
  \bibinfo{author}{L.~Leal}, \bibinfo{author}{J.~P. Lestone},
  \bibinfo{author}{C.~Lubitz}, \bibinfo{author}{J.~I. M{\'a}rquez~Dami{\'a}n},
  \bibinfo{author}{C.~M. Mattoon}, \bibinfo{author}{E.~A. McCutchan},
  \bibinfo{author}{S.~Mughabghab}, \bibinfo{author}{P.~Navratil},
  \bibinfo{author}{D.~Neudecker}, \bibinfo{author}{G.~P.~A. Nobre},
  \bibinfo{author}{G.~Noguere}, \bibinfo{author}{M.~Paris},
  \bibinfo{author}{M.~T. Pigni}, \bibinfo{author}{A.~J. Plompen},
  \bibinfo{author}{B.~Pritychenko}, \bibinfo{author}{V.~G. Pronyaev},
  \bibinfo{author}{D.~Roubtsov}, \bibinfo{author}{D.~Rochman},
  \bibinfo{author}{P.~Romano}, \bibinfo{author}{P.~Schillebeeckx},
  \bibinfo{author}{S.~Simakov}, \bibinfo{author}{M.~Sin},
  \bibinfo{author}{I.~Sirakov}, \bibinfo{author}{B.~Sleaford},
  \bibinfo{author}{V.~Sobes}, \bibinfo{author}{E.~S. Soukhovitskii},
  \bibinfo{author}{I.~Stetcu}, \bibinfo{author}{P.~Talou},
  \bibinfo{author}{I.~Thompson}, \bibinfo{author}{S.~van~der Marck},
  \bibinfo{author}{L.~Welser-Sherrill}, \bibinfo{author}{D.~Wiarda},
  \bibinfo{author}{M.~White}, \bibinfo{author}{J.~L. Wormald},
  \bibinfo{author}{R.~Q. Wright}, \bibinfo{author}{M.~Zerkle},
  \bibinfo{author}{G.~\v{Z}erovnik}, \bibinfo{author}{Y.~Zhu},
\newblock \bibinfo{title}{{ENDF/B-VIII.0}: The 8th major release of the nuclear
  reaction data library with {CIELO}-project cross sections, new standards and
  thermal scattering data},
\newblock \bibinfo{journal}{Nuclear Data Sheets} \bibinfo{volume}{148}
  (\bibinfo{year}{2018}) \bibinfo{pages}{1--142}.
  \DOIprefix\doi{10.1016/j.nds.2018.02.001}.
\bibitem[{Plompen et~al.(2020)Plompen, Cabellos, De~Saint~Jean, Fleming,
  Algora, Angelone, Archier, Bauge, Bersillon, Blokhin, Cantargi, Chebboubi,
  Diez, Duarte, Dupont, Dyrda, Erasmus, Fiorito, Fischer, Flammini, Foligno,
  Gilbert, Granada, Haeck, Hambsch, Helgesson, Hilaire, Hill, Hursin, Ichou,
  Jacqmin, Jansky, Jouanne, Kellett, Kim, Kim, Kodeli, Koning, Konobeyev,
  Kopecky, Kos, Kr{\'a}sa, Leal, Leclaire, Leconte, Lee, Leeb, Litaize,
  Majerle, M{\'a}rquez~Dami{\'a}n, Michel-Sendis, Mills, Morillon, Nogu{\`e}re,
  Pecchia, Pelloni, Pereslavtsev, Perry, Rochman, R\"{o}hrmoser, Romain,
  Romojaro, Roubtsov, Sauvan, Schillebeeckx, Schmidt, Serot, Simakov, Sirakov,
  Sj\"{o}strand, Stankovskiy, Sublet, Tamagno, Trkov, van~der Marck,
  {\'A}lvarez-Velarde, Villari, Ware, Yokoyama, and
  \v{Z}erovnik}]{Plompen2020Joint}
\bibinfo{author}{A.~J.~M. Plompen}, \bibinfo{author}{O.~Cabellos},
  \bibinfo{author}{C.~De~Saint~Jean}, \bibinfo{author}{M.~Fleming},
  \bibinfo{author}{A.~Algora}, \bibinfo{author}{M.~Angelone},
  \bibinfo{author}{P.~Archier}, \bibinfo{author}{E.~Bauge},
  \bibinfo{author}{O.~Bersillon}, \bibinfo{author}{A.~Blokhin},
  \bibinfo{author}{F.~Cantargi}, \bibinfo{author}{A.~Chebboubi},
  \bibinfo{author}{C.~Diez}, \bibinfo{author}{H.~Duarte},
  \bibinfo{author}{E.~Dupont}, \bibinfo{author}{J.~Dyrda},
  \bibinfo{author}{B.~Erasmus}, \bibinfo{author}{L.~Fiorito},
  \bibinfo{author}{U.~Fischer}, \bibinfo{author}{D.~Flammini},
  \bibinfo{author}{D.~Foligno}, \bibinfo{author}{M.~R. Gilbert},
  \bibinfo{author}{J.~R. Granada}, \bibinfo{author}{W.~Haeck},
  \bibinfo{author}{F.-J. Hambsch}, \bibinfo{author}{P.~Helgesson},
  \bibinfo{author}{S.~Hilaire}, \bibinfo{author}{I.~Hill},
  \bibinfo{author}{M.~Hursin}, \bibinfo{author}{R.~Ichou},
  \bibinfo{author}{R.~Jacqmin}, \bibinfo{author}{B.~Jansky},
  \bibinfo{author}{C.~Jouanne}, \bibinfo{author}{M.~A. Kellett},
  \bibinfo{author}{D.~H. Kim}, \bibinfo{author}{H.~I. Kim},
  \bibinfo{author}{I.~Kodeli}, \bibinfo{author}{A.~J. Koning},
  \bibinfo{author}{A.~Y. Konobeyev}, \bibinfo{author}{S.~Kopecky},
  \bibinfo{author}{B.~Kos}, \bibinfo{author}{A.~Kr{\'a}sa},
  \bibinfo{author}{L.~C. Leal}, \bibinfo{author}{N.~Leclaire},
  \bibinfo{author}{P.~Leconte}, \bibinfo{author}{Y.~O. Lee},
  \bibinfo{author}{H.~Leeb}, \bibinfo{author}{O.~Litaize},
  \bibinfo{author}{M.~Majerle}, \bibinfo{author}{J.~I. M{\'a}rquez~Dami{\'a}n},
  \bibinfo{author}{F.~Michel-Sendis}, \bibinfo{author}{R.~W. Mills},
  \bibinfo{author}{B.~Morillon}, \bibinfo{author}{G.~Nogu{\`e}re},
  \bibinfo{author}{M.~Pecchia}, \bibinfo{author}{S.~Pelloni},
  \bibinfo{author}{P.~Pereslavtsev}, \bibinfo{author}{R.~J. Perry},
  \bibinfo{author}{D.~Rochman}, \bibinfo{author}{A.~R\"{o}hrmoser},
  \bibinfo{author}{P.~Romain}, \bibinfo{author}{P.~Romojaro},
  \bibinfo{author}{D.~Roubtsov}, \bibinfo{author}{P.~Sauvan},
  \bibinfo{author}{P.~Schillebeeckx}, \bibinfo{author}{K.~H. Schmidt},
  \bibinfo{author}{O.~Serot}, \bibinfo{author}{S.~Simakov},
  \bibinfo{author}{I.~Sirakov}, \bibinfo{author}{H.~Sj\"{o}strand},
  \bibinfo{author}{A.~Stankovskiy}, \bibinfo{author}{J.~C. Sublet},
  \bibinfo{author}{P.~Tamagno}, \bibinfo{author}{A.~Trkov},
  \bibinfo{author}{S.~van~der Marck}, \bibinfo{author}{F.~{\'A}lvarez-Velarde},
  \bibinfo{author}{R.~Villari}, \bibinfo{author}{T.~C. Ware},
  \bibinfo{author}{K.~Yokoyama}, \bibinfo{author}{G.~\v{Z}erovnik},
\newblock \bibinfo{title}{The joint evaluated fission and fusion nuclear data
  library, {JEFF-3.3}},
\newblock \bibinfo{journal}{The European Physical Journal A}
  \bibinfo{volume}{56} (\bibinfo{year}{2020}) \bibinfo{pages}{181}.
  \DOIprefix\doi{10.1140/epja/s10050-020-00141-9}.
\bibitem[{Iwamoto et~al.(2023)Iwamoto, Iwamoto, Kunieda, Minato, Nakayama, Abe,
  Tsubakihara, Okumura, Ishizuka, Yoshida, Chiba, Otuka, Sublet, Iwamoto,
  Yamamoto, Nagaya, Tada, Konno, Matsuda, Yokoyama, Taninaka, Oizumi,
  Fukushima, Okita, Chiba, Sato, Ohta, and Kwon}]{Iwamoto2023Japanese}
\bibinfo{author}{O.~Iwamoto}, \bibinfo{author}{N.~Iwamoto},
  \bibinfo{author}{S.~Kunieda}, \bibinfo{author}{F.~Minato},
  \bibinfo{author}{S.~Nakayama}, \bibinfo{author}{Y.~Abe},
  \bibinfo{author}{K.~Tsubakihara}, \bibinfo{author}{S.~Okumura},
  \bibinfo{author}{C.~Ishizuka}, \bibinfo{author}{T.~Yoshida},
  \bibinfo{author}{S.~Chiba}, \bibinfo{author}{N.~Otuka},
  \bibinfo{author}{J.-C. Sublet}, \bibinfo{author}{H.~Iwamoto},
  \bibinfo{author}{K.~Yamamoto}, \bibinfo{author}{Y.~Nagaya},
  \bibinfo{author}{K.~Tada}, \bibinfo{author}{C.~Konno},
  \bibinfo{author}{N.~Matsuda}, \bibinfo{author}{K.~Yokoyama},
  \bibinfo{author}{H.~Taninaka}, \bibinfo{author}{A.~Oizumi},
  \bibinfo{author}{M.~Fukushima}, \bibinfo{author}{S.~Okita},
  \bibinfo{author}{G.~Chiba}, \bibinfo{author}{S.~Sato},
  \bibinfo{author}{M.~Ohta}, \bibinfo{author}{S.~Kwon},
\newblock \bibinfo{title}{Japanese evaluated nuclear data library version 5:
  {JENDL-5}},
\newblock \bibinfo{journal}{Journal of Nuclear Science and Technology}
  \bibinfo{volume}{60} (\bibinfo{year}{2023}) \bibinfo{pages}{1--60}.
  \DOIprefix\doi{10.1080/00223131.2022.2141903}.
\bibitem[{Kondev et~al.(2021)Kondev, Wang, Huang, Naimi, and
  Audi}]{Kondev2021NUBASE2020}
\bibinfo{author}{F.~G. Kondev}, \bibinfo{author}{M.~Wang},
  \bibinfo{author}{W.~J. Huang}, \bibinfo{author}{S.~Naimi},
  \bibinfo{author}{G.~Audi},
\newblock \bibinfo{title}{The {NUBASE2020} evaluation of nuclear physics
  properties},
\newblock \bibinfo{journal}{Chinese Physics C} \bibinfo{volume}{45}
  (\bibinfo{year}{2021}) \bibinfo{pages}{030001}.
  \DOIprefix\doi{10.1088/1674-1137/abddae}.
\bibitem[{Hopkins and Diven(1962)}]{Hopkins1962Neutron}
\bibinfo{author}{J.~C. Hopkins}, \bibinfo{author}{B.~C. Diven},
\newblock \bibinfo{title}{Neutron capture to fission ratios in {U}$^{233}$,
  {U}$^{235}$, {Pu}$^{239}$},
\newblock \bibinfo{journal}{Nuclear Science and Engineering}
  \bibinfo{volume}{12} (\bibinfo{year}{1962}) \bibinfo{pages}{169--177}.
  \DOIprefix\doi{10.13182/NSE62-A26055}, \bibinfo{note}{{EXFOR} 12331}.
\bibitem[{Otuka et~al.(2014)Otuka, Dupont, Semkova, Pritychenko, Blokhin,
  Aikawa, Babykina, Bossant, Chen, Dunaeva, Forrest, Fukahori, Furutachi,
  Ganesan, Ge, Gritzay, Herman, Hlava\v{c}, Kat\={o}, Lalremruata, Lee,
  Makinaga, Matsumoto, Mikhaylyukova, Pikulina, Pronyaev, Saxena, Schwerer,
  Simakov, Soppera, Suzuki, Tak{\'a}cs, Tao, Taova, T{\'a}rk{\'a}nyi, Varlamov,
  Wang, Yang, Zerkin, and Zhuang}]{Otuka2014Towards}
\bibinfo{author}{N.~Otuka}, \bibinfo{author}{E.~Dupont},
  \bibinfo{author}{V.~Semkova}, \bibinfo{author}{B.~Pritychenko},
  \bibinfo{author}{A.~I. Blokhin}, \bibinfo{author}{M.~Aikawa},
  \bibinfo{author}{S.~Babykina}, \bibinfo{author}{M.~Bossant},
  \bibinfo{author}{G.~Chen}, \bibinfo{author}{S.~Dunaeva},
  \bibinfo{author}{R.~A. Forrest}, \bibinfo{author}{T.~Fukahori},
  \bibinfo{author}{N.~Furutachi}, \bibinfo{author}{S.~Ganesan},
  \bibinfo{author}{Z.~Ge}, \bibinfo{author}{O.~O. Gritzay},
  \bibinfo{author}{M.~Herman}, \bibinfo{author}{S.~Hlava\v{c}},
  \bibinfo{author}{K.~Kat\={o}}, \bibinfo{author}{B.~Lalremruata},
  \bibinfo{author}{Y.~Lee}, \bibinfo{author}{A.~Makinaga},
  \bibinfo{author}{K.~Matsumoto}, \bibinfo{author}{M.~Mikhaylyukova},
  \bibinfo{author}{G.~Pikulina}, \bibinfo{author}{V.~Pronyaev},
  \bibinfo{author}{A.~Saxena}, \bibinfo{author}{O.~Schwerer},
  \bibinfo{author}{S.~Simakov}, \bibinfo{author}{N.~Soppera},
  \bibinfo{author}{R.~Suzuki}, \bibinfo{author}{S.~Tak{\'a}cs},
  \bibinfo{author}{X.~Tao}, \bibinfo{author}{S.~Taova},
  \bibinfo{author}{F.~T{\'a}rk{\'a}nyi}, \bibinfo{author}{V.~Varlamov},
  \bibinfo{author}{J.~Wang}, \bibinfo{author}{S.~Yang},
  \bibinfo{author}{V.~Zerkin}, \bibinfo{author}{Y.~Zhuang},
\newblock \bibinfo{title}{Towards a more complete and accurate experimental
  nuclear reaction data library ({{EXFOR}}): International collaboration
  between {N}uclear {R}eaction {D}ata {C}entres ({NRDC})},
\newblock \bibinfo{journal}{Nuclear Data Sheets} \bibinfo{volume}{120}
  (\bibinfo{year}{2014}) \bibinfo{pages}{272--276}.
  \DOIprefix\doi{10.1016/j.nds.2014.07.065}.
\bibitem[{Leal-Cidoncha et~al.(2023)Leal-Cidoncha, Couture, Bond, Bredeweg,
  Fry, Kawano, Lovell, Rusev, Stetcu, Ullmann, Leal, and
  Pigni}]{Leal-Cidoncha2023Measurement}
\bibinfo{author}{E.~Leal-Cidoncha}, \bibinfo{author}{A.~Couture},
  \bibinfo{author}{E.~M. Bond}, \bibinfo{author}{T.~A. Bredeweg},
  \bibinfo{author}{C.~Fry}, \bibinfo{author}{T.~Kawano}, \bibinfo{author}{A.~E.
  Lovell}, \bibinfo{author}{G.~Rusev}, \bibinfo{author}{I.~Stetcu},
  \bibinfo{author}{J.~L. Ullmann}, \bibinfo{author}{L.~Leal},
  \bibinfo{author}{M.~T. Pigni},
\newblock \bibinfo{title}{Measurement of the neutron-induced capture-to-fission
  cross section ratio in $^{233}${U} at {LANSCE}},
\newblock \bibinfo{journal}{Physical Review C} \bibinfo{volume}{108}
  (\bibinfo{year}{2023}) \bibinfo{pages}{014608}.
  \DOIprefix\doi{10.1103/PhysRevC.108.014608}, \bibinfo{note}{{EXFOR} 14819}.
\bibitem[{Otuka and Iwamoto(2022)}]{Otuka2022aEXFOR}
\bibinfo{author}{N.~Otuka}, \bibinfo{author}{O.~Iwamoto},
\newblock \bibinfo{title}{{{EXFOR}}-based simultaneous evaluation of
  neutron-induced uranium and plutonium fission cross sections for {JENDL-5}},
\newblock \bibinfo{journal}{Journal of Nuclear Science and Technology}
  \bibinfo{volume}{59} (\bibinfo{year}{2022}) \bibinfo{pages}{1004--1036}.
  \DOIprefix\doi{10.1080/00223131.2022.2030259}.
\bibitem[{Tarr{\'i}o et~al.(2023)Tarr{\'i}o, Tassan-Got, Duran, Leong,
  Paradela, Audouin, Leal-Cidoncha, Le~Naour, Caama{\~n}o, Ventura, Altstadt,
  Andrzejewski, Barbagallo, B{\'e}cares, Be\v{c}v{\'a}\v{r}, Belloni,
  Berthoumieux, Billowes, Boccone, Bosnar, Brugger, Calviani, Calvi{\~n}o,
  Cano-Ott, Carrapi\c{c}o, Cerutti, Chiaveri, Chin, Colonna, Cort{\'e}s,
  Cort{\'e}s-Giraldo, Diakaki, Domingo-Pardo, Dzysiuk, Eleftheriadis, Ferrari,
  Fraval, Ganesan, Garc{\'i}a, Giubrone, G{\'o}mez-Hornillos, Gon\c{c}alves,
  Gonz{\'a}lez-Romero, Griesmayer, Guerrero, Gunsing, Gurusamy, Jenkins,
  Jericha, Kadi, K\"{a}ppeler, Karadimos, Koehler, Kokkoris, Krti\v{c}ka,
  Kroll, Langer, Lederer, Leeb, Losito, Manousos, Marganiec, Mart{\'i}nez,
  Massimi, Mastinu, Mastromarco, Meaze, Mendoza, Mengoni, Milazzo, Mingrone,
  Mirea, Mondalaers, Pavlik, Perkowski, Plompen, Praena, Quesada, Rauscher,
  Reifarth, Riego, Robles, Roman, Rubbia, Sarmento, Schillebeeckx, Schmidt,
  Tagliente, Tain, Tsinganis, Valenta, Vannini, Variale, Vaz, Versaci,
  Vermeulen, Vlachoudis, Vlastou, Wallner, Ware, Weigand, Wei{\ss}, Wright, and
  \v{Z}ugec}]{Tarrio2023Neutron}
\bibinfo{author}{D.~Tarr{\'i}o}, \bibinfo{author}{L.~Tassan-Got},
  \bibinfo{author}{I.~Duran}, \bibinfo{author}{L.~S. Leong},
  \bibinfo{author}{C.~Paradela}, \bibinfo{author}{L.~Audouin},
  \bibinfo{author}{E.~Leal-Cidoncha}, \bibinfo{author}{C.~Le~Naour},
  \bibinfo{author}{M.~Caama{\~n}o}, \bibinfo{author}{A.~Ventura},
  \bibinfo{author}{S.~Altstadt}, \bibinfo{author}{J.~Andrzejewski},
  \bibinfo{author}{M.~Barbagallo}, \bibinfo{author}{V.~B{\'e}cares},
  \bibinfo{author}{F.~Be\v{c}v{\'a}\v{r}}, \bibinfo{author}{F.~Belloni},
  \bibinfo{author}{E.~Berthoumieux}, \bibinfo{author}{J.~Billowes},
  \bibinfo{author}{V.~Boccone}, \bibinfo{author}{D.~Bosnar},
  \bibinfo{author}{M.~Brugger}, \bibinfo{author}{M.~Calviani},
  \bibinfo{author}{F.~Calvi{\~n}o}, \bibinfo{author}{D.~Cano-Ott},
  \bibinfo{author}{C.~Carrapi\c{c}o}, \bibinfo{author}{F.~Cerutti},
  \bibinfo{author}{E.~Chiaveri}, \bibinfo{author}{M.~Chin},
  \bibinfo{author}{N.~Colonna}, \bibinfo{author}{G.~Cort{\'e}s},
  \bibinfo{author}{M.~A. Cort{\'e}s-Giraldo}, \bibinfo{author}{M.~Diakaki},
  \bibinfo{author}{C.~Domingo-Pardo}, \bibinfo{author}{N.~Dzysiuk},
  \bibinfo{author}{C.~Eleftheriadis}, \bibinfo{author}{A.~Ferrari},
  \bibinfo{author}{K.~Fraval}, \bibinfo{author}{S.~Ganesan},
  \bibinfo{author}{A.~R. Garc{\'i}a}, \bibinfo{author}{G.~Giubrone},
  \bibinfo{author}{M.~B. G{\'o}mez-Hornillos}, \bibinfo{author}{I.~F.
  Gon\c{c}alves}, \bibinfo{author}{E.~Gonz{\'a}lez-Romero},
  \bibinfo{author}{E.~Griesmayer}, \bibinfo{author}{C.~Guerrero},
  \bibinfo{author}{F.~Gunsing}, \bibinfo{author}{P.~Gurusamy},
  \bibinfo{author}{D.~G. Jenkins}, \bibinfo{author}{E.~Jericha},
  \bibinfo{author}{Y.~Kadi}, \bibinfo{author}{F.~K\"{a}ppeler},
  \bibinfo{author}{D.~Karadimos}, \bibinfo{author}{P.~Koehler},
  \bibinfo{author}{M.~Kokkoris}, \bibinfo{author}{M.~Krti\v{c}ka},
  \bibinfo{author}{J.~Kroll}, \bibinfo{author}{C.~Langer},
  \bibinfo{author}{C.~Lederer}, \bibinfo{author}{H.~Leeb},
  \bibinfo{author}{R.~Losito}, \bibinfo{author}{A.~Manousos},
  \bibinfo{author}{J.~Marganiec}, \bibinfo{author}{T.~Mart{\'i}nez},
  \bibinfo{author}{C.~Massimi}, \bibinfo{author}{P.~F. Mastinu},
  \bibinfo{author}{M.~Mastromarco}, \bibinfo{author}{M.~Meaze},
  \bibinfo{author}{E.~Mendoza}, \bibinfo{author}{A.~Mengoni},
  \bibinfo{author}{P.~M. Milazzo}, \bibinfo{author}{F.~Mingrone},
  \bibinfo{author}{M.~Mirea}, \bibinfo{author}{W.~Mondalaers},
  \bibinfo{author}{A.~Pavlik}, \bibinfo{author}{J.~Perkowski},
  \bibinfo{author}{A.~Plompen}, \bibinfo{author}{J.~Praena},
  \bibinfo{author}{J.~M. Quesada}, \bibinfo{author}{T.~Rauscher},
  \bibinfo{author}{R.~Reifarth}, \bibinfo{author}{A.~Riego},
  \bibinfo{author}{M.~S. Robles}, \bibinfo{author}{F.~Roman},
  \bibinfo{author}{C.~Rubbia}, \bibinfo{author}{R.~Sarmento},
  \bibinfo{author}{P.~Schillebeeckx}, \bibinfo{author}{S.~Schmidt},
  \bibinfo{author}{G.~Tagliente}, \bibinfo{author}{J.~L. Tain},
  \bibinfo{author}{A.~Tsinganis}, \bibinfo{author}{S.~Valenta},
  \bibinfo{author}{G.~Vannini}, \bibinfo{author}{V.~Variale},
  \bibinfo{author}{P.~Vaz}, \bibinfo{author}{R.~Versaci},
  \bibinfo{author}{M.~J. Vermeulen}, \bibinfo{author}{V.~Vlachoudis},
  \bibinfo{author}{R.~Vlastou}, \bibinfo{author}{A.~Wallner},
  \bibinfo{author}{T.~Ware}, \bibinfo{author}{M.~Weigand},
  \bibinfo{author}{C.~Wei{\ss}}, \bibinfo{author}{T.~J. Wright},
  \bibinfo{author}{P.~\v{Z}ugec},
\newblock \bibinfo{title}{Neutron-induced fission cross sections of
  $^{232}${Th} and $^{233}${U} up to 1~{GeV} using parallel plate avalanche
  counters at the {CERN} {n\_TOF} facility},
\newblock \bibinfo{journal}{Physical Review C} \bibinfo{volume}{107}
  (\bibinfo{year}{2023}) \bibinfo{pages}{044616}.
  \DOIprefix\doi{10.1103/PhysRevC.107.044616}, \bibinfo{note}{{EXFOR} 23654}.
\bibitem[{Devi et~al.(2024)Devi, Otuka, and Ganesan}]{Devi2024EXFOR}
\bibinfo{author}{V.~Devi}, \bibinfo{author}{N.~Otuka},
  \bibinfo{author}{S.~Ganesan},
\newblock \bibinfo{title}{{{EXFOR}}-based simultaneous evaluation for fast
  neutron-induced fission cross section of thorium-232},
\newblock \bibinfo{journal}{Journal of Nuclear Science and Technology}
  \bibinfo{volume}{61} (\bibinfo{year}{2024}) \bibinfo{pages}{44--56}.
  \DOIprefix\doi{10.1080/00223131.2023.2256715}.
\bibitem[{Okuyama et~al.(2024)Okuyama, Otuka, Chiba, and
  Iwamoto}]{Okuyama2024EXFOR}
\bibinfo{author}{R.~Okuyama}, \bibinfo{author}{N.~Otuka},
  \bibinfo{author}{G.~Chiba}, \bibinfo{author}{O.~Iwamoto},
\newblock \bibinfo{title}{{{EXFOR}}-based simultaneous evaluation for
  neutron-induced fission cross section of plutonium-242},
\newblock \bibinfo{journal}{Journal of Nuclear Science and Technology}
  \bibinfo{volume}{61} (\bibinfo{year}{2024}) \bibinfo{pages}{57--67}.
  \DOIprefix\doi{10.1080/00223131.2023.2267070}.
\bibitem[{Iwamoto(2007)}]{Iwamoto2007Development}
\bibinfo{author}{O.~Iwamoto},
\newblock \bibinfo{title}{Development of a comprehensive code for nuclear data
  evaluation, {CCONE}, and validation using neutron-induced cross sections for
  uranium isotopes},
\newblock \bibinfo{journal}{Journal of Nuclear Science and Technology}
  \bibinfo{volume}{44} (\bibinfo{year}{2007}) \bibinfo{pages}{687--697}.
  \DOIprefix\doi{10.1080/18811248.2007.9711857}.
\bibitem[{Shibata et~al.(2011)Shibata, Iwamoto, Nakagawa, Iwamoto, Ichihara,
  Kunieda, Chiba, Furutaka, Otuka, Ohsawa, Murata, Matsunobu, Zukeran, Kamada,
  and Katakura}]{Shibata2011Japanese}
\bibinfo{author}{K.~Shibata}, \bibinfo{author}{O.~Iwamoto},
  \bibinfo{author}{T.~Nakagawa}, \bibinfo{author}{N.~Iwamoto},
  \bibinfo{author}{A.~Ichihara}, \bibinfo{author}{S.~Kunieda},
  \bibinfo{author}{S.~Chiba}, \bibinfo{author}{K.~Furutaka},
  \bibinfo{author}{N.~Otuka}, \bibinfo{author}{T.~Ohsawa},
  \bibinfo{author}{T.~Murata}, \bibinfo{author}{H.~Matsunobu},
  \bibinfo{author}{A.~Zukeran}, \bibinfo{author}{S.~Kamada},
  \bibinfo{author}{J.-I. Katakura},
\newblock \bibinfo{title}{{JENDL}-4.0: {A} new library for nuclear science and
  engineering},
\newblock \bibinfo{journal}{Journal of Nuclear Science and Technology}
  \bibinfo{volume}{48} (\bibinfo{year}{2011}) \bibinfo{pages}{1--30}.
  \DOIprefix\doi{10.1080/18811248.2011.9711675}.
\bibitem[{Young et~al.(1998)Young, Arthur, and
  Chadwick}]{Young1998Comprehensive}
\bibinfo{author}{P.~G. Young}, \bibinfo{author}{E.~D. Arthur},
  \bibinfo{author}{M.~B. Chadwick},
\newblock \bibinfo{title}{Comprehensive nuclear model calculations: Theory and
  use of the {GNASH} code},
\newblock in: \bibinfo{booktitle}{Proceedings of the Workshop on Nuclear
  Reaction Data and Nuclear Reactors: Physics, Design, and Safety, Trieste, 15
  April--17 May 1996}, volume~\bibinfo{volume}{1}, \bibinfo{year}{1998}, pp.
  \bibinfo{pages}{227--404}.
\bibitem[{Chadwick et~al.(2006)Chadwick, Oblo\v{z}insk{\'y}, Herman, Greene,
  McKnight, Smith, Young, MacFarlane, Hale, Frankle, Kahler, Kawano, Little,
  Madland, Moller, Mosteller, Page, Talou, Trellue, White, Wilson, Arcilla,
  Dunford, Mughabghab, Pritychenko, Rochman, Sonzogni, Lubitz, Trumbull,
  Weinman, Brown, Cullen, Heinrichs, McNabb, Derrien, Dunn, Larson, Leal,
  Carlson, Block, Briggs, Cheng, Huria, Zerkle, Kozier, Courcelle, Pronyaev,
  and van~der Marck}]{Chadwick2006ENDF}
\bibinfo{author}{M.~B. Chadwick}, \bibinfo{author}{P.~Oblo\v{z}insk{\'y}},
  \bibinfo{author}{M.~Herman}, \bibinfo{author}{N.~M. Greene},
  \bibinfo{author}{R.~D. McKnight}, \bibinfo{author}{D.~L. Smith},
  \bibinfo{author}{P.~G. Young}, \bibinfo{author}{R.~E. MacFarlane},
  \bibinfo{author}{G.~M. Hale}, \bibinfo{author}{S.~C. Frankle},
  \bibinfo{author}{A.~C. Kahler}, \bibinfo{author}{T.~Kawano},
  \bibinfo{author}{R.~C. Little}, \bibinfo{author}{D.~G. Madland},
  \bibinfo{author}{P.~Moller}, \bibinfo{author}{R.~D. Mosteller},
  \bibinfo{author}{P.~R. Page}, \bibinfo{author}{P.~Talou},
  \bibinfo{author}{H.~Trellue}, \bibinfo{author}{M.~C. White},
  \bibinfo{author}{W.~B. Wilson}, \bibinfo{author}{R.~Arcilla},
  \bibinfo{author}{C.~L. Dunford}, \bibinfo{author}{S.~F. Mughabghab},
  \bibinfo{author}{B.~Pritychenko}, \bibinfo{author}{D.~Rochman},
  \bibinfo{author}{A.~A. Sonzogni}, \bibinfo{author}{C.~R. Lubitz},
  \bibinfo{author}{T.~H. Trumbull}, \bibinfo{author}{J.~P. Weinman},
  \bibinfo{author}{D.~A. Brown}, \bibinfo{author}{D.~E. Cullen},
  \bibinfo{author}{D.~P. Heinrichs}, \bibinfo{author}{D.~P. McNabb},
  \bibinfo{author}{H.~Derrien}, \bibinfo{author}{M.~E. Dunn},
  \bibinfo{author}{N.~M. Larson}, \bibinfo{author}{L.~C. Leal},
  \bibinfo{author}{A.~D. Carlson}, \bibinfo{author}{R.~C. Block},
  \bibinfo{author}{J.~B. Briggs}, \bibinfo{author}{E.~T. Cheng},
  \bibinfo{author}{H.~C. Huria}, \bibinfo{author}{M.~L. Zerkle},
  \bibinfo{author}{K.~S. Kozier}, \bibinfo{author}{A.~Courcelle},
  \bibinfo{author}{V.~Pronyaev}, \bibinfo{author}{S.~C. van~der Marck},
\newblock \bibinfo{title}{{ENDF/B-VII.0}: {N}ext generation evaluated nuclear
  data library for nuclear science and technology},
\newblock \bibinfo{journal}{Nuclear Data Sheets} \bibinfo{volume}{107}
  (\bibinfo{year}{2006}) \bibinfo{pages}{2931--3060}.
  \DOIprefix\doi{10.1016/j.nds.2006.11.001}.
\bibitem[{Chadwick et~al.(2011)Chadwick, Herman, Oblo\v{z}insk{\'y}, Dunn,
  Danon, Kahler, Smith, Pritychenko, Arbanas, Arcilla, Brewer, Brown, Capote,
  Carlson, Cho, Derrien, Guber, Hale, Hoblit, Holloway, Johnson, Kawano,
  Kiedrowski, Kim, Kunieda, Larson, Leal, Lestone, Little, McCutchan,
  MacFarlane, MacInnes, Mattoon, McKnight, Mughabghab, Nobre, Palmiotti,
  Palumbo, Pigni, Pronyaev, Sayer, Sonzogni, Summers, Talou, Thompson, Trkov,
  Vogt, van~der Marck, Wallner, White, Wiarda, and Young}]{Chadwick2011ENDF}
\bibinfo{author}{M.~B. Chadwick}, \bibinfo{author}{M.~Herman},
  \bibinfo{author}{P.~Oblo\v{z}insk{\'y}}, \bibinfo{author}{M.~E. Dunn},
  \bibinfo{author}{Y.~Danon}, \bibinfo{author}{A.~C. Kahler},
  \bibinfo{author}{D.~L. Smith}, \bibinfo{author}{B.~Pritychenko},
  \bibinfo{author}{G.~Arbanas}, \bibinfo{author}{R.~Arcilla},
  \bibinfo{author}{R.~Brewer}, \bibinfo{author}{D.~A. Brown},
  \bibinfo{author}{R.~Capote}, \bibinfo{author}{A.~D. Carlson},
  \bibinfo{author}{Y.~S. Cho}, \bibinfo{author}{H.~Derrien},
  \bibinfo{author}{K.~Guber}, \bibinfo{author}{G.~M. Hale},
  \bibinfo{author}{S.~Hoblit}, \bibinfo{author}{S.~Holloway},
  \bibinfo{author}{T.~D. Johnson}, \bibinfo{author}{T.~Kawano},
  \bibinfo{author}{B.~C. Kiedrowski}, \bibinfo{author}{H.~Kim},
  \bibinfo{author}{S.~Kunieda}, \bibinfo{author}{N.~M. Larson},
  \bibinfo{author}{L.~Leal}, \bibinfo{author}{J.~P. Lestone},
  \bibinfo{author}{R.~C. Little}, \bibinfo{author}{E.~A. McCutchan},
  \bibinfo{author}{R.~E. MacFarlane}, \bibinfo{author}{M.~MacInnes},
  \bibinfo{author}{C.~M. Mattoon}, \bibinfo{author}{R.~D. McKnight},
  \bibinfo{author}{S.~F. Mughabghab}, \bibinfo{author}{G.~P.~A. Nobre},
  \bibinfo{author}{G.~Palmiotti}, \bibinfo{author}{A.~Palumbo},
  \bibinfo{author}{M.~T. Pigni}, \bibinfo{author}{V.~G. Pronyaev},
  \bibinfo{author}{R.~O. Sayer}, \bibinfo{author}{A.~A. Sonzogni},
  \bibinfo{author}{N.~C. Summers}, \bibinfo{author}{P.~Talou},
  \bibinfo{author}{I.~J. Thompson}, \bibinfo{author}{A.~Trkov},
  \bibinfo{author}{R.~L. Vogt}, \bibinfo{author}{S.~C. van~der Marck},
  \bibinfo{author}{A.~Wallner}, \bibinfo{author}{M.~C. White},
  \bibinfo{author}{D.~Wiarda}, \bibinfo{author}{P.~G. Young},
\newblock \bibinfo{title}{{ENDF/B-VII.1} nuclear data for science and
  technology: Cross sections, covariances, fission product yields and decay
  data},
\newblock \bibinfo{journal}{Nuclear Data Sheets} \bibinfo{volume}{112}
  (\bibinfo{year}{2011}) \bibinfo{pages}{2887--2996}.
  \DOIprefix\doi{10.1016/j.nds.2011.11.002}.
\bibitem[{Yu et~al.(1999)Yu, Shen, and Cai}]{Yu1999Evaluation}
\bibinfo{author}{B.~Yu}, \bibinfo{author}{Q.~Shen}, \bibinfo{author}{C.~Cai},
  \bibinfo{title}{Evaluation of complete neutron nuclear data for $^{233}${U}},
  \bibinfo{type}{Technical Report} \bibinfo{number}{INDC(CPR)-049},
  International Atomic Energy Agency, \bibinfo{year}{1999}.
\bibitem[{Zhang(2006)}]{Zhang2006User}
\bibinfo{author}{J.~Zhang}, \bibinfo{title}{User manual of {FUNF} code for
  fissile material data calculation}, \bibinfo{type}{Technical Report}
  \bibinfo{number}{CNDC-0037}, China Institute of Atomic Energy,
  \bibinfo{year}{2006}.
\bibitem[{Kawano et~al.(2000{\natexlab{a}})Kawano, Matsunobu, Murata, Zukeran,
  Nakajima, Kawai, Iwamoto, Shibata, Nakagawa, Osawa, Baba, and
  Yoshida}]{Kawano2000evaluation}
\bibinfo{author}{T.~Kawano}, \bibinfo{author}{H.~Matsunobu},
  \bibinfo{author}{T.~Murata}, \bibinfo{author}{A.~Zukeran},
  \bibinfo{author}{Y.~Nakajima}, \bibinfo{author}{M.~Kawai},
  \bibinfo{author}{O.~Iwamoto}, \bibinfo{author}{K.~Shibata},
  \bibinfo{author}{T.~Nakagawa}, \bibinfo{author}{T.~Osawa},
  \bibinfo{author}{M.~Baba}, \bibinfo{author}{T.~Yoshida},
  \bibinfo{title}{Evaluation of fission cross sections and covariances for
  $^{233}${U}, $^{235}${U}, $^{238}${U}, $^{239}${Pu}, $^{240}${Pu}, and
  $^{241}${Pu}}, \bibinfo{type}{Technical Report}
  \bibinfo{number}{JAERI-Research 2000-004}, Japan Atomic Energy Research
  Institute, \bibinfo{year}{2000}{\natexlab{a}}.
  \DOIprefix\doi{10.11484/jaeri-research-2000-004}.
\bibitem[{Kawano et~al.(2000{\natexlab{b}})Kawano, Matsunobu, Murata, Zukeran,
  Nakajima, Kawai, Iwamoto, Shibata, Nakagawa, Ohsawa, Baba, and
  Yoshida}]{Kawano2000Simultaneous}
\bibinfo{author}{T.~Kawano}, \bibinfo{author}{H.~Matsunobu},
  \bibinfo{author}{T.~Murata}, \bibinfo{author}{A.~Zukeran},
  \bibinfo{author}{Y.~Nakajima}, \bibinfo{author}{M.~Kawai},
  \bibinfo{author}{O.~Iwamoto}, \bibinfo{author}{K.~Shibata},
  \bibinfo{author}{T.~Nakagawa}, \bibinfo{author}{T.~Ohsawa},
  \bibinfo{author}{M.~Baba}, \bibinfo{author}{T.~Yoshida},
\newblock \bibinfo{title}{Simultaneous evaluation of fission cross sections of
  uranium and plutonium isotopes for {JENDL}-3.3},
\newblock \bibinfo{journal}{Journal of Nuclear Science and Technology}
  \bibinfo{volume}{37} (\bibinfo{year}{2000}{\natexlab{b}})
  \bibinfo{pages}{327--334}. \DOIprefix\doi{10.1080/18811248.2000.9714902}.
\bibitem[{OEC(2007)}]{OECD2007CINDA}
\bibinfo{title}{{CINDA 2006: The Comprehensive Index of Nuclear Reaction Data:
  Archive 1935-2006}}, \bibinfo{publisher}{Nuclear Energy Agency, Organisation
  for Economic Co-operation and Development}, \bibinfo{year}{2007}.
\bibitem[{Otuka and Iwamoto(2022)}]{Otuka2022bEXFOR}
\bibinfo{author}{N.~Otuka}, \bibinfo{author}{O.~Iwamoto},
  \bibinfo{title}{{{EXFOR}}-based simultaneous evaluation of neutron-induced
  uranium and plutonium fission cross sections for {JENDL-5}: {I}nputs and
  outputs}, \bibinfo{type}{Technical Report} \bibinfo{number}{JAEA-Data/Code
  2022-005}, Japan Atomic Energy Agency, \bibinfo{year}{2022}.
  \DOIprefix\doi{10.11484/jaea-data-code-2022-005}.
\bibitem[{Arlt et~al.(1981)Arlt, Josch, Musiol, Ortlepp, Pausch, Richter,
  Teichner, and Wagner}]{Arlt1981Absolute}
\bibinfo{author}{R.~Arlt}, \bibinfo{author}{M.~Josch},
  \bibinfo{author}{G.~Musiol}, \bibinfo{author}{H.-G. Ortlepp},
  \bibinfo{author}{G.~Pausch}, \bibinfo{author}{U.~Richter},
  \bibinfo{author}{R.~Teichner}, \bibinfo{author}{W.~Wagner},
\newblock \bibinfo{title}{Absolute fission cross section measurement on
  $^{235}${U} at 8.4~{MeV} neutron energy},
\newblock in: \bibinfo{booktitle}{Proceedings of the X-th International
  Symposium on Selected Topics of the Interaction of Fast Neutrons and Heavy
  Ions with Atomic Nuclei, Gaussig, 17--21 November 1980},
  \bibinfo{number}{INDC(GDR)-19}, \bibinfo{publisher}{International Atomic
  Energy Agency}, \bibinfo{year}{1981}, pp. \bibinfo{pages}{35--39}.
  \bibinfo{note}{{EXFOR} 31833}.
\bibitem[{Otuka and Iwamoto(2023)}]{Otuka2023Simultaneous}
\bibinfo{author}{N.~Otuka}, \bibinfo{author}{O.~Iwamoto},
\newblock \bibinfo{title}{Simultaneous evaluation of uranium and plutonium fast
  neutron fission cross sections up to 200~{MeV} for {JENDL-5} and its
  updates},
\newblock \bibinfo{journal}{EPJ Web of Conferences} \bibinfo{volume}{284}
  (\bibinfo{year}{2023}) \bibinfo{pages}{08011}.
  \DOIprefix\doi{10.1051/epjconf/202328408011}.
\bibitem[{Duran et~al.(2019)Duran, Paradela, Caama{\~n}o, Cabanelas,
  Tassan-Got, and Audouin}]{Duran2019High}
\bibinfo{author}{I.~Duran}, \bibinfo{author}{C.~Paradela},
  \bibinfo{author}{M.~Caama{\~n}o}, \bibinfo{author}{P.~Cabanelas},
  \bibinfo{author}{L.~Tassan-Got}, \bibinfo{author}{L.~Audouin},
\newblock \bibinfo{title}{High-resolution evaluation of the {U5}(n,f) cross
  section from 3~{keV} to 30~{keV}},
\newblock \bibinfo{journal}{EPJ Web of Conferences} \bibinfo{volume}{211}
  (\bibinfo{year}{2019}) \bibinfo{pages}{02003}.
  \DOIprefix\doi{10.1051/epjconf/201921102003}, \bibinfo{note}{{EXFOR} 23294}.
\bibitem[{Belloni et~al.(2022)Belloni, Eykens, Heyse, Matei, Moens, Nolte,
  Plompen, Richter, Sibbens, Vanleeuw, and Wynants}]{Belloni2022Neutron}
\bibinfo{author}{F.~Belloni}, \bibinfo{author}{R.~Eykens},
  \bibinfo{author}{J.~Heyse}, \bibinfo{author}{C.~Matei},
  \bibinfo{author}{A.~Moens}, \bibinfo{author}{R.~Nolte},
  \bibinfo{author}{A.~J.~M. Plompen}, \bibinfo{author}{S.~Richter},
  \bibinfo{author}{G.~Sibbens}, \bibinfo{author}{D.~Vanleeuw},
  \bibinfo{author}{R.~Wynants},
\newblock \bibinfo{title}{Neutron induced fission cross section measurements of
  $^{240}${Pu} and $^{242}${Pu} relative to the neutron--proton scattering
  cross section at 2.5 and 14.8~{MeV}},
\newblock \bibinfo{journal}{The European Physical Journal A}
  \bibinfo{volume}{58} (\bibinfo{year}{2022}) \bibinfo{pages}{227}.
  \DOIprefix\doi{10.1140/epja/s10050-022-00858-9}, \bibinfo{note}{{EXFOR}
  23653}.
\bibitem[{Ren et~al.(2023)Ren, Yang, Liu, Ye, Wen, Wen, Guo, Chen, Yi, Sun,
  Yan, Han, Liu, Chen, Ye, Bai, An, Bai, Bao, Cao, Cheng, Cui, Fan, Feng, Gu,
  Guo, Han, He, He, He, Huang, Huang, Huang, Ji, Ji, Jiang, Jiang, Jing, Kang,
  Kang, Li, Li, Li, Li, Li, Li, Liu, Luan, Ma, Ning, Qi, Ren, Ruan, Song, Sun,
  Sun, Sun, Tan, Tang, Tang, Wang, Wang, Wang, Wang, Wang, Wang, Wu, Wu, Wu,
  Xie, Yu, Yu, Yu, Zhang, Zhang, Zhang, Zhang, Zhang, Zhang, Zhang, Zhang,
  Zhang, Zhao, Zhou, Zhou, Zhu, Zhu, and Zhu}]{Ren2023Measurement}
\bibinfo{author}{Z.~Ren}, \bibinfo{author}{Y.~Yang}, \bibinfo{author}{R.~Liu},
  \bibinfo{author}{B.~Ye}, \bibinfo{author}{Z.~Wen}, \bibinfo{author}{J.~Wen},
  \bibinfo{author}{H.~Guo}, \bibinfo{author}{Y.~Chen}, \bibinfo{author}{H.~Yi},
  \bibinfo{author}{W.~Sun}, \bibinfo{author}{J.~Yan}, \bibinfo{author}{Z.~Han},
  \bibinfo{author}{X.~Liu}, \bibinfo{author}{Q.~Chen}, \bibinfo{author}{T.~Ye},
  \bibinfo{author}{J.~Bai}, \bibinfo{author}{Q.~An}, \bibinfo{author}{H.~Bai},
  \bibinfo{author}{J.~Bao}, \bibinfo{author}{P.~Cao},
  \bibinfo{author}{P.~Cheng}, \bibinfo{author}{Z.~Cui},
  \bibinfo{author}{R.~Fan}, \bibinfo{author}{C.~Feng}, \bibinfo{author}{M.~Gu},
  \bibinfo{author}{F.~Guo}, \bibinfo{author}{C.~Han}, \bibinfo{author}{G.~He},
  \bibinfo{author}{Y.~He}, \bibinfo{author}{Y.~He}, \bibinfo{author}{H.~Huang},
  \bibinfo{author}{W.~Huang}, \bibinfo{author}{X.~Huang},
  \bibinfo{author}{X.~Ji}, \bibinfo{author}{X.~Ji}, \bibinfo{author}{H.~Jiang},
  \bibinfo{author}{W.~Jiang}, \bibinfo{author}{H.~Jing},
  \bibinfo{author}{L.~Kang}, \bibinfo{author}{M.~Kang},
  \bibinfo{author}{B.~Li}, \bibinfo{author}{L.~Li}, \bibinfo{author}{Q.~Li},
  \bibinfo{author}{X.~Li}, \bibinfo{author}{Y.~Li}, \bibinfo{author}{Y.~Li},
  \bibinfo{author}{S.~Liu}, \bibinfo{author}{G.~Luan}, \bibinfo{author}{Y.~Ma},
  \bibinfo{author}{C.~Ning}, \bibinfo{author}{B.~Qi}, \bibinfo{author}{J.~Ren},
  \bibinfo{author}{X.~Ruan}, \bibinfo{author}{Z.~Song},
  \bibinfo{author}{H.~Sun}, \bibinfo{author}{X.~Sun}, \bibinfo{author}{Z.~Sun},
  \bibinfo{author}{Z.~Tan}, \bibinfo{author}{H.~Tang},
  \bibinfo{author}{J.~Tang}, \bibinfo{author}{P.~Wang},
  \bibinfo{author}{Q.~Wang}, \bibinfo{author}{T.~Wang},
  \bibinfo{author}{Y.~Wang}, \bibinfo{author}{Z.~Wang},
  \bibinfo{author}{Z.~Wang}, \bibinfo{author}{Q.~Wu}, \bibinfo{author}{X.~Wu},
  \bibinfo{author}{X.~Wu}, \bibinfo{author}{L.~Xie}, \bibinfo{author}{L.~Yu},
  \bibinfo{author}{T.~Yu}, \bibinfo{author}{Y.~Yu}, \bibinfo{author}{G.~Zhang},
  \bibinfo{author}{J.~Zhang}, \bibinfo{author}{L.~Zhang},
  \bibinfo{author}{L.~Zhang}, \bibinfo{author}{Q.~Zhang},
  \bibinfo{author}{Q.~Zhang}, \bibinfo{author}{X.~Zhang},
  \bibinfo{author}{Y.~Zhang}, \bibinfo{author}{Z.~Zhang},
  \bibinfo{author}{Y.~Zhao}, \bibinfo{author}{L.~Zhou},
  \bibinfo{author}{Z.~Zhou}, \bibinfo{author}{D.~Zhu},
  \bibinfo{author}{K.~Zhu}, \bibinfo{author}{P.~Zhu},
\newblock \bibinfo{title}{Measurement of the $^{236,238}${U}(n,f) cross
  sections from the threshold to 200~{MeV} at {CSNS Back-n}},
\newblock \bibinfo{journal}{The European Physical Journal A}
  \bibinfo{volume}{59} (\bibinfo{year}{2023}) \bibinfo{pages}{5}.
  \DOIprefix\doi{10.1140/epja/s10050-022-00910-8}, \bibinfo{note}{{EXFOR}
  32886}.
\bibitem[{Michalopoulou et~al.(2023)Michalopoulou, Stamatopoulos, Diakaki,
  Tsinganis, Vlastou, Kokkoris, Patronis, Eleme, Macina, Tassan-Got, Colonna,
  Chiaveri, Ventura, Schillebeeckx, Heyse, Sibbens, Alaerts, Borella, Moens,
  Vanleeuw, Aberle, Alcayne, Amaducci, Andrzejewski, Audouin, Babiano-Suarez,
  Bacak, Barbagallo, Bennett, Berthoumieux, Billowes, Bosnar, Brown, Busso,
  Caama{\~n}o, Caballero, Calvi{\~n}o, Calviani, Cano-Ott, Casanovas, Cerutti,
  Cort{\'e}s, Cort{\'e}s-Giraldo, Cosentino, Cristallo, Damone, Davies, Dietz,
  Domingo-Pardo, Dressler, Ducasse, Dupont, Dur{\'a}n,
  Fern{\'a}ndez-Dom{\'i}nguez, Ferrari, Finocchiaro, Furman, G\"{o}bel, Garg,
  Gawlik-Rami\k{e}ga, Gilardoni, Gon\c{c}alves, Gonz{\'a}lez-Romero, Guerrero,
  Gunsing, Harada, Heinitz, Jenkins, Junghans, K\"{a}ppeler, Kadi, Kimura,
  Knapov{\'a}, Kopatch, Krti\v{c}ka, Kurtulgil, Ladarescu, Lederer-Woods, Leeb,
  Lerendegui-Marco, Lonsdale, Manna, Mart{\'i}nez, Masi, Massimi, Mastinu,
  Mastromarco, Maugeri, Mazzone, Mendoza, Mengoni, Milazzo, Mingrone,
  Moreno-Soto, Musumarra, Negret, Nolte, Og{\'a}llar, Oprea, Pavlik, Perkowski,
  Petrone, Piersanti, Pirovano, Porras, Praena, Quesada, Ramos-Doval, Rauscher,
  Reifarth, Rochman, Romanets, Rubbia, Sabat{\'e}-Gilarte, Saxena, Schumann,
  Sekhar, Smith, Sosnin, Sprung, Tagliente, Tain, Tarife{\~n}o-Saldivia,
  Thomas, Torres-S{\'a}nchez, Ulrich, Urlass, Valenta, Vannini, Variale, Vaz,
  Vescovi, Vlachoudis, Wallner, Woods, Wright, and
  \v{Z}ugec}]{Michalopoulou2023Measurement}
\bibinfo{author}{V.~Michalopoulou}, \bibinfo{author}{A.~Stamatopoulos},
  \bibinfo{author}{M.~Diakaki}, \bibinfo{author}{A.~Tsinganis},
  \bibinfo{author}{R.~Vlastou}, \bibinfo{author}{M.~Kokkoris},
  \bibinfo{author}{N.~Patronis}, \bibinfo{author}{Z.~Eleme},
  \bibinfo{author}{D.~Macina}, \bibinfo{author}{L.~Tassan-Got},
  \bibinfo{author}{N.~Colonna}, \bibinfo{author}{E.~Chiaveri},
  \bibinfo{author}{A.~Ventura}, \bibinfo{author}{P.~Schillebeeckx},
  \bibinfo{author}{J.~Heyse}, \bibinfo{author}{G.~Sibbens},
  \bibinfo{author}{G.~Alaerts}, \bibinfo{author}{A.~Borella},
  \bibinfo{author}{A.~Moens}, \bibinfo{author}{D.~Vanleeuw},
  \bibinfo{author}{O.~Aberle}, \bibinfo{author}{V.~Alcayne},
  \bibinfo{author}{S.~Amaducci}, \bibinfo{author}{J.~Andrzejewski},
  \bibinfo{author}{L.~Audouin}, \bibinfo{author}{V.~Babiano-Suarez},
  \bibinfo{author}{M.~Bacak}, \bibinfo{author}{M.~Barbagallo},
  \bibinfo{author}{S.~Bennett}, \bibinfo{author}{E.~Berthoumieux},
  \bibinfo{author}{J.~Billowes}, \bibinfo{author}{D.~Bosnar},
  \bibinfo{author}{A.~Brown}, \bibinfo{author}{M.~Busso},
  \bibinfo{author}{M.~Caama{\~n}o}, \bibinfo{author}{L.~Caballero},
  \bibinfo{author}{F.~Calvi{\~n}o}, \bibinfo{author}{M.~Calviani},
  \bibinfo{author}{D.~Cano-Ott}, \bibinfo{author}{A.~Casanovas},
  \bibinfo{author}{F.~Cerutti}, \bibinfo{author}{G.~Cort{\'e}s},
  \bibinfo{author}{M.~A. Cort{\'e}s-Giraldo}, \bibinfo{author}{L.~Cosentino},
  \bibinfo{author}{S.~Cristallo}, \bibinfo{author}{L.~A. Damone},
  \bibinfo{author}{P.~J. Davies}, \bibinfo{author}{M.~Dietz},
  \bibinfo{author}{C.~Domingo-Pardo}, \bibinfo{author}{R.~Dressler},
  \bibinfo{author}{Q.~Ducasse}, \bibinfo{author}{E.~Dupont},
  \bibinfo{author}{I.~Dur{\'a}n},
  \bibinfo{author}{B.~Fern{\'a}ndez-Dom{\'i}nguez},
  \bibinfo{author}{A.~Ferrari}, \bibinfo{author}{P.~Finocchiaro},
  \bibinfo{author}{V.~Furman}, \bibinfo{author}{K.~G\"{o}bel},
  \bibinfo{author}{R.~Garg}, \bibinfo{author}{A.~Gawlik-Rami\k{e}ga},
  \bibinfo{author}{S.~Gilardoni}, \bibinfo{author}{I.~F. Gon\c{c}alves},
  \bibinfo{author}{E.~Gonz{\'a}lez-Romero}, \bibinfo{author}{C.~Guerrero},
  \bibinfo{author}{F.~Gunsing}, \bibinfo{author}{H.~Harada},
  \bibinfo{author}{S.~Heinitz}, \bibinfo{author}{D.~G. Jenkins},
  \bibinfo{author}{A.~Junghans}, \bibinfo{author}{F.~K\"{a}ppeler},
  \bibinfo{author}{Y.~Kadi}, \bibinfo{author}{A.~Kimura},
  \bibinfo{author}{I.~Knapov{\'a}}, \bibinfo{author}{Y.~Kopatch},
  \bibinfo{author}{M.~Krti\v{c}ka}, \bibinfo{author}{D.~Kurtulgil},
  \bibinfo{author}{I.~Ladarescu}, \bibinfo{author}{C.~Lederer-Woods},
  \bibinfo{author}{H.~Leeb}, \bibinfo{author}{J.~Lerendegui-Marco},
  \bibinfo{author}{S.~J. Lonsdale}, \bibinfo{author}{A.~Manna},
  \bibinfo{author}{T.~Mart{\'i}nez}, \bibinfo{author}{A.~Masi},
  \bibinfo{author}{C.~Massimi}, \bibinfo{author}{P.~Mastinu},
  \bibinfo{author}{M.~Mastromarco}, \bibinfo{author}{E.~A. Maugeri},
  \bibinfo{author}{A.~Mazzone}, \bibinfo{author}{E.~Mendoza},
  \bibinfo{author}{A.~Mengoni}, \bibinfo{author}{P.~M. Milazzo},
  \bibinfo{author}{F.~Mingrone}, \bibinfo{author}{J.~Moreno-Soto},
  \bibinfo{author}{A.~Musumarra}, \bibinfo{author}{A.~Negret},
  \bibinfo{author}{R.~Nolte}, \bibinfo{author}{F.~Og{\'a}llar},
  \bibinfo{author}{A.~Oprea}, \bibinfo{author}{A.~Pavlik},
  \bibinfo{author}{J.~Perkowski}, \bibinfo{author}{C.~Petrone},
  \bibinfo{author}{L.~Piersanti}, \bibinfo{author}{E.~Pirovano},
  \bibinfo{author}{I.~Porras}, \bibinfo{author}{J.~Praena},
  \bibinfo{author}{J.~M. Quesada}, \bibinfo{author}{D.~Ramos-Doval},
  \bibinfo{author}{T.~Rauscher}, \bibinfo{author}{R.~Reifarth},
  \bibinfo{author}{D.~Rochman}, \bibinfo{author}{Y.~Romanets},
  \bibinfo{author}{C.~Rubbia}, \bibinfo{author}{M.~Sabat{\'e}-Gilarte},
  \bibinfo{author}{A.~Saxena}, \bibinfo{author}{D.~Schumann},
  \bibinfo{author}{A.~Sekhar}, \bibinfo{author}{A.~G. Smith},
  \bibinfo{author}{N.~V. Sosnin}, \bibinfo{author}{P.~Sprung},
  \bibinfo{author}{G.~Tagliente}, \bibinfo{author}{J.~L. Tain},
  \bibinfo{author}{A.~Tarife{\~n}o-Saldivia}, \bibinfo{author}{T.~Thomas},
  \bibinfo{author}{P.~Torres-S{\'a}nchez}, \bibinfo{author}{J.~Ulrich},
  \bibinfo{author}{S.~Urlass}, \bibinfo{author}{S.~Valenta},
  \bibinfo{author}{G.~Vannini}, \bibinfo{author}{V.~Variale},
  \bibinfo{author}{P.~Vaz}, \bibinfo{author}{D.~Vescovi},
  \bibinfo{author}{V.~Vlachoudis}, \bibinfo{author}{A.~Wallner},
  \bibinfo{author}{P.~J. Woods}, \bibinfo{author}{T.~Wright},
  \bibinfo{author}{P.~\v{Z}ugec},
\newblock \bibinfo{title}{Measurement of the neutron-induced fission cross
  section of $^{230}${Th} at the {CERN} n\_{TOF} facility},
\newblock \bibinfo{journal}{Physical Review C} \bibinfo{volume}{108}
  (\bibinfo{year}{2023}) \bibinfo{pages}{014616}.
  \DOIprefix\doi{10.1103/PhysRevC.108.014616}, \bibinfo{note}{{EXFOR} 23657}.
\bibitem[{Vorobyev et~al.(2023)Vorobyev, Gagarski, Shcherbakov, Vaishnene, and
  Barabanov}]{Vorobyev2023Measurement}
\bibinfo{author}{A.~S. Vorobyev}, \bibinfo{author}{A.~M. Gagarski},
  \bibinfo{author}{O.~A. Shcherbakov}, \bibinfo{author}{L.~A. Vaishnene},
  \bibinfo{author}{A.~L. Barabanov},
\newblock \bibinfo{title}{Measurement of the cross section for the
  neutron-induced fission of $^{238}${U} nuclei in the energy range of 0.3--500
  {MeV}},
\newblock \bibinfo{journal}{JETP Letters} \bibinfo{volume}{117}
  (\bibinfo{year}{2023}) \bibinfo{pages}{557--565}.
  \DOIprefix\doi{10.1134/S0021364023600787}, \bibinfo{note}{{EXFOR} 41756}.
\bibitem[{Snyder et~al.(2021)Snyder, Anastasiou, Bowden, Bundgaard, Casperson,
  Cebra, Classen, Dongwi, Fotiades, Gearhart, Geppert-Kleinrath, Greife,
  Hagmann, Heffner, Hensle, Higgins, Isenhower, Kazkaz, Kemnitz, King, Klay,
  Latta, Leal-Cidoncha, Loveland, Magee, Manning, Mendenhall, Monterial, Mosby,
  Neudecker, Prokop, Sangiorgio, Schmitt, Seilhan, Tovesson, Towell, Walsh,
  Watson, Yao, and Younes}]{Snyder2021Measurement}
\bibinfo{author}{L.~Snyder}, \bibinfo{author}{M.~Anastasiou},
  \bibinfo{author}{N.~S. Bowden}, \bibinfo{author}{J.~Bundgaard},
  \bibinfo{author}{R.~J. Casperson}, \bibinfo{author}{D.~A. Cebra},
  \bibinfo{author}{T.~Classen}, \bibinfo{author}{D.~H. Dongwi},
  \bibinfo{author}{N.~Fotiades}, \bibinfo{author}{J.~Gearhart},
  \bibinfo{author}{V.~Geppert-Kleinrath}, \bibinfo{author}{U.~Greife},
  \bibinfo{author}{C.~Hagmann}, \bibinfo{author}{M.~Heffner},
  \bibinfo{author}{D.~Hensle}, \bibinfo{author}{D.~Higgins},
  \bibinfo{author}{L.~D. Isenhower}, \bibinfo{author}{K.~Kazkaz},
  \bibinfo{author}{A.~Kemnitz}, \bibinfo{author}{J.~King},
  \bibinfo{author}{J.~L. Klay}, \bibinfo{author}{J.~Latta},
  \bibinfo{author}{E.~Leal-Cidoncha}, \bibinfo{author}{W.~Loveland},
  \bibinfo{author}{J.~A. Magee}, \bibinfo{author}{B.~Manning},
  \bibinfo{author}{M.~P. Mendenhall}, \bibinfo{author}{M.~Monterial},
  \bibinfo{author}{S.~Mosby}, \bibinfo{author}{D.~Neudecker},
  \bibinfo{author}{C.~Prokop}, \bibinfo{author}{S.~Sangiorgio},
  \bibinfo{author}{K.~T. Schmitt}, \bibinfo{author}{B.~Seilhan},
  \bibinfo{author}{F.~Tovesson}, \bibinfo{author}{R.~S. Towell},
  \bibinfo{author}{N.~Walsh}, \bibinfo{author}{T.~S. Watson},
  \bibinfo{author}{L.~Yao}, \bibinfo{author}{W.~Younes},
\newblock \bibinfo{title}{Measurement of the $^{239}${Pu}(n,f)/$^{235}${U}(n,f)
  cross-section ratio with the {NIFFTE} fission {Time Projection Chamber}},
\newblock \bibinfo{journal}{Nuclear Data Sheets} \bibinfo{volume}{178}
  (\bibinfo{year}{2021}) \bibinfo{pages}{1--40}.
  \DOIprefix\doi{10.1016/j.nds.2021.11.001}, \bibinfo{note}{{EXFOR} 14721}.
\bibitem[{Schmittroth and Schenter(1980)}]{Schmittroth1980Finite}
\bibinfo{author}{F.~Schmittroth}, \bibinfo{author}{R.~E. Schenter},
\newblock \bibinfo{title}{Finite element basis in data adjustment},
\newblock \bibinfo{journal}{Nuclear Science and Engineering}
  \bibinfo{volume}{74} (\bibinfo{year}{1980}) \bibinfo{pages}{168--177}.
  \DOIprefix\doi{10.13182/NSE80-A20116}.
\bibitem[{Mannhart(2013)}]{Mannhart2013Small}
\bibinfo{author}{W.~Mannhart}, \bibinfo{title}{A small guide to generating
  covariances of experimental data}, \bibinfo{type}{Technical Report}
  \bibinfo{number}{INDC(NDS)-0588 Rev.}, International Atomic Energy Agency,
  \bibinfo{year}{2013}.
\bibitem[{Hirtz et~al.(2024)Hirtz, David, Cugnon, Leya,
  Rodr{\'i}guez-S{\'a}nchez, and Schnabel}]{Hirtz2024Parameter}
\bibinfo{author}{J.~Hirtz}, \bibinfo{author}{J.~C. David},
  \bibinfo{author}{J.~Cugnon}, \bibinfo{author}{I.~Leya},
  \bibinfo{author}{J.~L. Rodr{\'i}guez-S{\'a}nchez},
  \bibinfo{author}{G.~Schnabel},
\newblock \bibinfo{title}{Parameter optimisation using bayesian inference for
  spallation models},
\newblock \bibinfo{journal}{The European Physical Journal A}
  \bibinfo{volume}{60} (\bibinfo{year}{2024}) \bibinfo{pages}{149}.
  \DOIprefix\doi{10.1140/epja/s10050-024-01370-y}.
\bibitem[{Pigni et~al.(2023)Pigni, Trkov, and Capote}]{Pigni2023INDEN}
\bibinfo{author}{M.~Pigni}, \bibinfo{author}{A.~Trkov},
  \bibinfo{author}{A.~Capote}, \bibinfo{howpublished}{ENDF-6 file Ver.
  u233e81b2\_R4nu5fc\_ENDF downloaded from https://nds.iaea.org/INDEN/},
  \bibinfo{year}{2023}.
\bibitem[{Poenitz and Aumeier(1997)}]{Poenitz1997Simultaneous}
\bibinfo{author}{W.~P. Poenitz}, \bibinfo{author}{S.~E. Aumeier},
  \bibinfo{title}{The simultaneous evaluation of the standards and other cross
  sections of importance for technology}, \bibinfo{type}{Technical Report}
  \bibinfo{number}{ANL/NDM-139}, Argonne National Laboratory,
  \bibinfo{year}{1997}.
\bibitem[{Guber et~al.(2000)Guber, Spencer, Leal, Harvey, Hill, Dos~Santos,
  Sayer, and Larson}]{Guber2000New}
\bibinfo{author}{K.~H. Guber}, \bibinfo{author}{R.~R. Spencer},
  \bibinfo{author}{L.~C. Leal}, \bibinfo{author}{J.~A. Harvey},
  \bibinfo{author}{N.~W. Hill}, \bibinfo{author}{G.~Dos~Santos},
  \bibinfo{author}{R.~O. Sayer}, \bibinfo{author}{D.~C. Larson},
\newblock \bibinfo{title}{New high-resolution fission cross-section
  measurements of $^{233}${U} in the 0.4-{eV} to 700-{keV} energy range},
\newblock \bibinfo{journal}{Nuclear Science and Engineering}
  \bibinfo{volume}{135} (\bibinfo{year}{2000}) \bibinfo{pages}{141--149}.
  \DOIprefix\doi{10.13182/NSE00-A2130}, \bibinfo{note}{{EXFOR} 13890}.
\bibitem[{Blons et~al.(1971)Blons, Derrien, and
  Michaudon}]{Blons1971MeasurementAnalysis}
\bibinfo{author}{J.~Blons}, \bibinfo{author}{H.~Derrien},
  \bibinfo{author}{A.~Michaudon},
\newblock \bibinfo{title}{Measurement and analysis of the fission cross
  sections of $^{233}${U} and $^{235}${U} for neutron energies below 30~{keV}},
\newblock in: \bibinfo{booktitle}{Proceedings of the Third Conference on
  Neutron Cross Sections and Technology, Knoxville, 15--17 March 1971},
  volume~\bibinfo{volume}{2}, \bibinfo{year}{1971}, pp.
  \bibinfo{pages}{829--835}. \bibinfo{note}{{EXFOR} 20446,51008}.
\bibitem[{Nagaya et~al.(2017)Nagaya, Okumura, Sakurai, and
  Mori}]{Nagaya2017MVP}
\bibinfo{author}{Y.~Nagaya}, \bibinfo{author}{K.~Okumura},
  \bibinfo{author}{T.~Sakurai}, \bibinfo{author}{T.~Mori},
  \bibinfo{title}{{MVP/GMVP Version 3}: {G}eneral purpose {Monte Carlo} codes
  for neutron and photon transport calculations based on continuous energy and
  multigroup methods}, \bibinfo{type}{Technical Report}
  \bibinfo{number}{JAEA-Data/Code 2016-018}, Japan Atomic Energy Agency,
  \bibinfo{year}{2017}. \DOIprefix\doi{10.11484/jaea-data-code-2016-018}.
\bibitem[{Kuwagaki and Nagaya(2017)}]{Kuwagai2017Integral}
\bibinfo{author}{K.~Kuwagaki}, \bibinfo{author}{Y.~Nagaya},
  \bibinfo{title}{Integral benchmark test of {JENDL-4.0} for U-233 systems with
  {ICSBEP} {Handbook}}, \bibinfo{type}{Technical Report}
  \bibinfo{number}{JAEA-Data/Code 2017-007}, Japan Atomic Energy Agency,
  \bibinfo{year}{2017}. \DOIprefix\doi{10.11484/jaea-data-code-2017-007}.
\bibitem[{{JENDL Committee Reactor Integral Test Working
  Group}(2023)}]{JENDL2023Compilation}
\bibinfo{author}{{JENDL Committee Reactor Integral Test Working Group}},
  \bibinfo{title}{Compilation of the data book on light water reactor benchmark
  to develop {JENDL}; {U}tilization and extension of 2017 report
  ({JAEA-Data/Code 2017-006)}}, \bibinfo{type}{Technical Report}
  \bibinfo{number}{JAEA-Data/Code 2023-004}, Japan Atomic Energy Agency,
  \bibinfo{year}{2023}. \DOIprefix\doi{10.11484/jaea-data-code-2023-004}.
\bibitem[{OEC(2016)}]{OECD2014International}
\bibinfo{title}{{I}nternational {H}andbook of {E}valuated {C}riticality
  {S}afety {B}enchmark {E}xperiments}, \bibinfo{publisher}{Nuclear Energy
  Agency, Organisation for Economic Co-operation and Development},
  \bibinfo{year}{2016}.
\bibitem[{Tada et~al.(2024)Tada, Nagaya, Taninaka, Yokoyama, Okita, Oizumi,
  Fukushima, and Nakayama}]{Tada2024JENDL-5}
\bibinfo{author}{K.~Tada}, \bibinfo{author}{Y.~Nagaya},
  \bibinfo{author}{H.~Taninaka}, \bibinfo{author}{K.~Yokoyama},
  \bibinfo{author}{S.~Okita}, \bibinfo{author}{A.~Oizumi},
  \bibinfo{author}{M.~Fukushima}, \bibinfo{author}{S.~Nakayama},
\newblock \bibinfo{title}{{JENDL-5} benchmarking for fission reactor
  applications},
\newblock \bibinfo{journal}{Journal of Nuclear Science and Technology}
  \bibinfo{volume}{61} (\bibinfo{year}{2024}) \bibinfo{pages}{2--22}.
  \DOIprefix\doi{10.1080/00223131.2023.2197439}.
\bibitem[{Kawano(2019)}]{Kawano2019Dece}
\bibinfo{author}{T.~Kawano},
\newblock \bibinfo{title}{{DeCE}: the {ENDF}-6 data interface and nuclear data
  evaluation assist code},
\newblock \bibinfo{journal}{Journal of Nuclear Science and Technology}
  \bibinfo{volume}{56} (\bibinfo{year}{2019}) \bibinfo{pages}{1029--1035}.
  \DOIprefix\doi{10.1080/00223131.2019.1637797}.
\bibitem[{Mori et~al.(1996)Mori, Nakagawa, and Kaneko}]{Mori1996Neutron}
\bibinfo{author}{T.~Mori}, \bibinfo{author}{M.~Nakagawa},
  \bibinfo{author}{K.~Kaneko}, \bibinfo{title}{Neutron cross section library
  production code system for continuous energy/{M}onte {C}arlo code {MVP}:
  {LICEM}}, \bibinfo{type}{Technical Report} \bibinfo{number}{JAERI-Data/Code
  96-018}, Japan Atomic Energy Research Institute, \bibinfo{year}{1996}.
  \DOIprefix\doi{10.11484/jaeri-data-code-96-018}.

\end{thebibliography}

\end{document}